\documentclass[a4paper,11pt]{article}
\usepackage{jheppub} 

\usepackage{placeins}

\usepackage[utf8]{inputenc}

\usepackage{varioref}
\usepackage{amsmath,amsfonts,amsthm,mathrsfs}
\usepackage{enumerate}
\usepackage{fancyvrb}
\usepackage{verbatim}
\usepackage{wrapfig}
\usepackage{appendix}
\usepackage{amstext}
\usepackage{amssymb}
\usepackage{graphicx}
\usepackage{color}
\usepackage{varioref}
\usepackage{multirow,graphics}
\usepackage{epstopdf}
\usepackage{bbm}
\usepackage{slashed}
\usepackage{psfrag}
\usepackage{relsize}
\usepackage{bbm}
\usepackage{mathrsfs}
\usepackage{epsfig}

\usepackage{mathtools}

       \usepackage{dutchcal}
       \usepackage{dsfont}

\DeclareMathAlphabet{\mathpzc}{OT1}{pzc}{m}{it}

\usepackage{tikz}
\usepackage{subcaption}

\usepackage{mathtools}
\usepackage{bigints}

\usetikzlibrary{decorations}

\usepackage{amstext}
\usepackage{amssymb}
\usepackage{amsmath}
\usepackage{verbatim}

\newcommand{\bea}{\begin{eqnarray}}
\newcommand{\eea}{\end{eqnarray}}

\def\la{\label}

\def\tt{\top}

\def\gQ{{\mathpzc{Q}\hspace{.07em}}}

\def\a {\alpha}
\def\b {\beta}

\numberwithin{equation}{section}

\def\<{\langle}
\def\>{\rangle}

\makeatletter
\@ifundefined{usebibtex}{} {}
\makeatother

\usepackage{upgreek}

\newcommand\eps{\epsilon}

\usepackage{esint}
\usepackage{breqn}

\def\i{\textsl{\textsf{i}}}

\makeatletter
\newcommand{\raisemath}[1]{\mathpalette{\raisem@th{#1}}}
\newcommand{\raisem@th}[3]{\raisebox{#1}{$#2#3$}}
\makeatother

\newcommand{\alg}[1]{\mathfrak{#1}}
\def\gl{\alg{gl}}
\def\DP{\textsf{p}}
\def\hw{\uplambda}
\def\Qa{Q}
\def\mb{\textsl{\textsf{m}}}
\def\hb{\eta}
\def\zt{\updelta}

\def\qeg{\textsl{\textsf{Z}}}
\def\i{\textsl{\textsf{i}}}

\def\mon{{\bf \textsf{M}}}
\def\mT{\textsf{T}}
\def\BB{\textsf{B}  }
\def\rv{\upxi}
\def\Sym{{\alg S}}
\def\sgn{{\rm sgn}}
\def\Mt{\pmb{\mathbb M}}
\def\gg{\textsl{g}}
\def\xx{\nu}

\preprint{ZMP-HH/26-15}

\author[a]{Gleb Arutyunov,}
\author[b]{~Hrachya Babujian}
\author[a]{and\,~Minghao Gao}
\smallskip

\affiliation[a]{II. Institut f\"ur Theoretische Physik, Universit\"at Hamburg, Luruper Chaussee 149, 22761
Hamburg, Germany\\
Zentrum  f\"{u}r  Mathematische  Physik,  Universit\"{a}t  Hamburg,  Bundesstrasse  55,  20146  Hamburg,Germany
\\}
\affiliation[b]{A.I. Alikhanyan National Science Laboratory, Yerevan Physics Institute,\\
 2 Alikhanyan brothers str, Yerevan 0036, Armenia\\
Beijing Institute of Mathematical Sciences and Applications, Beijing, China
\\
\\
\noindent\textit{E-mails:}
\href{gleb.arutyunov@desy.de,}{gleb.arutyunov@desy.de,}
\href{hrachya.babujian@gmail.com,}{hrachya.babujian@gmail.com,}\\
$~$~~~~~~~~~~~~\!\href{minghao.gao@studium.uni-hamburg.de}{minghao.gao@studium.uni-hamburg.de}
}

\title{Bethe Ansatz without Nesting}

\abstract{
We develop a non-nested Bethe ansatz description of rational $\gl_\ell$ spin chains in the vector representation. Starting from the quantum spectral curve and the separation-of-variables framework, we derive closed systems of Bethe equations involving only the momentum-carrying Bethe roots. The construction is worked out explicitly for the $\gl_3$ and $\gl_4$ spin chains and then generalized to arbitrary rank.
A central result of this work is the identification of a recursive hierarchy associated with the fundamental transfer matrices. The hierarchy is generated by regularity conditions of the lower transfer matrices and closes through a universal rank-$\ell$ equation $\mathcal{R}_{\ell}=0$.
This equation replaces the final level of the conventional nested Bethe ansatz
and eliminates all auxiliary Bethe roots. 
Consequently, the complete spectral data of an eigenstate are encoded solely in the first Baxter polynomial $Q_{1}(u)$.
We further obtain explicit expressions for the eigenvalues of all fundamental transfer matrices in terms of the momentum-carrying roots alone. The resulting formulation provides a compact characterization of the spectrum of rational $\gl_\ell$ spin chains and reveals a direct connection between the quantum spectral curve, transfer-matrix fusion relations, and a truncated 
$Q$-system underlying the non-nested description. Finally, we investigate the quasi-classical (Gaudin) limit of the non-nested Bethe equations. For the $\gl_3$ spin chain, we show that the leading non-trivial contribution gives rise to Gaudin equations whose pole-free form naturally defines a scalar third-order $\gl_3$ oper.
}

\begin{document}

\hskip9cm

\maketitle


\newpage

\section{Introduction}

The Bethe ansatz remains one of the most powerful methods for solving quantum integrable models. Since Bethe's original solution of the Heisenberg spin chain, a variety of algebraic and analytic formulations have been developed, including the algebraic Bethe ansatz, Baxter's functional approach, and the method of Separation of Variables (SoV). While the spectrum of rank-one integrable models can often be described in terms of a single Baxter polynomial, the situation becomes considerably more involved for higher-rank systems. In the standard nested Bethe ansatz for rational $\gl_\ell$ spin chains, the eigenstates and eigenvalues are parameterized by several sets of Bethe roots associated with different nesting levels.

The necessity of introducing auxiliary Bethe roots is one of the main technical complications of the nested Bethe ansatz. Although these roots encode the internal structure of the eigenstates, physical observables are ultimately determined by the transfer-matrix eigenvalues and by the momentum-carrying roots of the first nesting level, at least for spin chains in the vector 
or symmetric representations. This naturally raises the question of whether the auxiliary roots can be eliminated altogether and whether the complete spectral problem can be reformulated in terms of a single set of variables.

A natural framework for addressing this question is provided by the method of Separation of Variables (SoV). Originally introduced by Sklyanin for a wide class of classical and quantum integrable systems \cite{Sklyanin:1987Gaudin, Sklyanin:1990FBA,Sklyanin:1992SL3,Sklyanin:1995SoV,Sklyanin:1993}, the SoV approach reformulates the spectral problem in terms of separated coordinates and associated wave functions. In recent years substantial progress has been achieved in constructing SoV bases for rational $\gl_{\ell}$-invariant spin chains \cite{Gromov:2016itr, Ryan:2018xyz,Ryan:2020rfq}. A particularly important development was the proposal of a new operator $\BB(u)$ in \cite{Gromov:2016itr}, whose repeated action on the reference state generates transfer-matrix eigenstates. In contrast to the conventional nested algebraic Bethe ansatz, where one introduces a hierarchy of creation
operators associated with different nesting levels, the construction of \cite{Gromov:2016itr} employs a single operator $\BB(u)$, thereby providing a
much more economical description of the spectrum. The same work conjectured the spectrum of the operator $\BB(u)$ and identified the
corresponding separated variables. This conjecture was subsequently established analytically in \cite{Ryan:2018xyz,Ryan:2020rfq}, where
explicit SoV bases were constructed for rational $\gl_{\ell}$ spin chains and the corresponding separated wave functions were derived in the so-called companion twist frame.

A complementary and conceptually related approach was developed in
Refs.~\cite{Maillet:2018bga,Maillet:2019ypa}. This
construction is based on generating the left eigenstates of $\BB(u)$ through the action of
suitably chosen transfer matrices, or powers thereof, on the unique vacuum state. It immediately leads
to a separation of variables in the coordinate representation of the wave
functions and provides a powerful characterization of the spectrum
without relying on the traditional nested Bethe ansatz framework.
Together, these developments establish SoV as a natural language for the
spectral analysis of higher-rank quantum spin chains. 

One of the central outcomes of the SoV approach is that transfer-matrix
eigenstates admit a factorized representation in separated variables,
while the spectral problem is encoded into finite-difference equations
for Baxter $Q$-functions. In particular, the eigenvalues of the operator
$\BB(u)$ provide the separated coordinates, and the corresponding wave
functions factorize into products of Baxter polynomials. 


An important feature of the SoV construction is the appearance of a quantum spectral curve. For rational $\gl_\ell$ spin chains this curve takes the form of a finite-order functional equation satisfied by the first Baxter polynomial $Q_1(u)$. In the standard Bethe ansatz language, the polynomial $Q_1(u)$ contains precisely the momentum-carrying Bethe roots,
 whereas the remaining Baxter polynomials $Q_2(u),\ldots,Q_{\ell-1}(u)$ correspond to auxiliary nesting levels. The quantum spectral curve therefore suggests the possibility of eliminating all auxiliary roots and describing the spectrum solely through $Q_1(u)$. A detailed account of the SoV framework and its applications to rational $\gl_{\ell}$ spin chains can be found 
 in \cite{Arutyunov2026}.

The purpose of this paper is to develop this idea systematically. 
Starting from the quantum spectral curve arising in the SoV framework, we obtain a compact system of non-nested Bethe equations 
involving only the roots of the Baxter polynomial $Q_1(u)$:
\bea
\la{Comp}
\mathcal{R}_{\ell}(v_k)=0\, , ~~~k=1,\ldots, \mb_1.
\eea
Equations (\ref{Comp}), 
whose explicit form is given in (\ref{conjecture}), completely characterize the transfer-matrix spectrum in terms of the momentum-carrying Bethe roots $v_k$. 
They arise from evaluating the closure condition of a recursively generated hierarchy of transfer-matrix functions at the zeroes of $Q_1(u)$.
We first derive them explicitly for the $\gl_3$ and $\gl_4$ spin chains and then formulate their extension to arbitrary rank.

We also investigate the quasi-classical (Gaudin) limit of the non-nested Bethe equations. In the $\gl_3$ case, we show that the first non-trivial contribution 
appears at the second order in the coupling constant and yields a system of Gaudin-type equations. These equations admit a natural interpretation as the pole-free conditions for a scalar third-order $\gl_3$ oper, 
establishing a direct connection between the non-nested Bethe equations and the oper formulation of the Gaudin model. Although this framework also involves the momentum-carrying Baxter polynomial $Q_1(u)$, it appears there together with the remaining Baxter polynomials of the nested Bethe ansatz. By contrast, the present approach eliminates all auxiliary polynomials and derives a closed system of equations involving $Q_1(u)$ alone.

For the $\gl_3$ spin chain, another set of non-nested equations for the momentum-carrying Bethe roots was derived in \cite{Liashyk:2018qfc} using 
determinant identities related to the modified Izergin determinant \cite{Belliard:2019}  and earlier work \cite{Gorsky:2013xba}. 
Although this construction can be extended to arbitrary rank, its precise relation to the complete nested Bethe equations remains to be fully understood.

The present construction follows a different route based on the quantum
spectral curve. It yields a hierarchy of non-nested Bethe equations
valid for arbitrary rank, provides a recursive reconstruction of the
transfer-matrix eigenvalues, and establishes a direct connection between the non-nested Bethe equations and scalar opers through their  Gaudin limit.

The resulting equations (\ref{Comp}) encode the complete spectral data of an eigenstate in terms of the principal roots alone. Once these roots are known, transfer-matrix eigenvalues can be reconstructed directly from the quantum spectral curve without introducing any auxiliary Baxter polynomials. In this sense, the formulation developed here may be viewed as a higher-rank analogue of the ordinary Bethe equations for rank-one systems.

At a more conceptual level, our construction reveals the emergence of a universal hierarchy of auxiliary functions generated recursively from the regularity conditions of the lower transfer matrices. For a spin chain of rank $\ell$, the hierarchy terminates at a single functional equation  $\mathcal{R}_\ell(u)=0$, 
whose evaluation at the zeroes of $Q_1(u)$ yields the non-nested Bethe equations (\ref{Comp}). 
In this picture, the conventional auxiliary Bethe roots are replaced by a finite hierarchy of rational functions whose closure is governed by the single functional equation $\mathcal{R}_\ell(u)=0$. The resulting structure may be viewed as a truncated $Q$-system whose last node is fixed by the quantum spectral curve and the regularity of the highest transfer matrix.

Besides providing a compact description of the spectrum, the non-nested formulation reveals a direct connection between the Bethe ansatz and the quantum spectral curve. It also offers a new perspective on the role of auxiliary roots in higher-rank integrable models: rather than being fundamental variables, they emerge as auxiliary quantities whose effect can be encoded entirely in a finite-order functional equation for $Q_1(u)$.

The paper is organized as follows. In section 2, we briefly review the nested Bethe ansatz and the quantum spectral curve for rational $\gl_\ell$ spin chains. Sections 3 and 4 are devoted to the derivation of the non-nested Bethe equations for the $\gl_3$ and $\gl_4$ chains, respectively. In section 5, we present the generalization to arbitrary rank. 
In section 6, we investigate the Gaudin limit and show that the resulting non-nested Gaudin equations admit a natural interpretation in terms of scalar opers. Finally, section~7 contains 
our conclusions and discusses possible directions for future research. Appendix A provides numerical evidence that the non-nested Bethe equations reproduce the same set of principal Bethe roots as the standard nested Bethe equations.

\enlargethispage{2\baselineskip}
\section{Bethe Ansatz and Quantum Spectral Curve}
\paragraph{\textsf{Nested Bethe equations.}}\hspace{-0.2cm}Consider an inhomogeneous $\gl_{\ell}$ spin chain of length $N$, where the local spin at the $j$th site transforms in the irreducible $\gl_{\ell}$-module
with highest weight $(\hw_j^{(1)},\ldots , \hw_{j}^{(\ell)})$.  To write the corresponding Bethe equations in the concise form, we need to define two quantities. 
First, we introduce the so-called vacuum  polynomials that encode all the information about the representation content of the spin chain. They are given by
\pagebreak[3]
\begin{equation}
\la{DP}
\DP_{\a}(u)=\prod_{j=1}^N \big(u-u_j-\hb\, \hw_{j}^{(\a)} \big)
\, ,~~~~\a=1,\ldots, \ell\, .
\end{equation}
Here $u$ is the spectral parameter, $u_j$ with $j=1,\ldots, N$ are inhomogeneities and $\eta$ is a coupling constant. 
For the chain with all spins in the vector representation we have the highest weight $(1,0,\ldots, 0)$ for all sites and the corresponding vacuum polynomials are 
 \begin{equation}
\la{DPv}
\DP_{1}(u)=\prod_{j=1}^N \big(u-u_j-\hb \big)\, , ~~~~~\DP_{\a}(u)=\prod_{j=1}^N \big(u-u_j\big)\, ,~~~~\a=2,\ldots, \ell\, .
\end{equation}
Next, 
we introduce $\ell-1$ {\it Baxter's $Q$-polynomials}
\bea
\la{Bax_pol_Q}
~~~~\Qa_{\a}(u)=\prod_{j=1}^{\mb_{\a}} \big(u-u_j^{(\a)}\big)\, , ~~~~~~~\a=1,\ldots, \ell-1,
\eea  
which incorporate $\ell-1$ sets of Bethe roots $\{u_j^{(\a)}\}$, each set contains $\mb_\a$ roots. We will call the roots $v_j:=u_j^{(1)}$ as principal, and the others as auxiliary.
We also introduce  $\Qa_0=\Qa_{\ell}=1$.

With these definitions, the nested Bethe equations for the inhomogeneous $\gl_{\ell}$ spin chains are 
\bea
\la{Bethe_concise}
\frac{\Qa_{\a-1}\big(u_j^{(\a)}\big)}{\Qa_{\a-1}\big(u_j^{(\a)}-\hb\big)}\frac{\Qa_{\a}\big(u_j^{(\a)}-\hb\big)}{\Qa_{\a}\big(u_j^{(\a)}+\hb\big)}\frac{\Qa_{\a+1}\big(u_j^{(\a)}+\hb\big)}{\Qa_{\a+1}\big(u_j^{(\a)}\big)}
=-\frac{\zt_\a}{\zt_{\a+1}}\frac{\DP_{\a}\big(u_j^{(\a)}\big)~}{\DP_{\a+1}\big(u_j^{(\a)}\big)}
\, , 
\eea
where $\a=1,\ldots, \ell-1$, and for each $\a$, the index $j$ ranges from $1$ to $\mb_{\a}$. The variables $\zt_{\a}$ on the right hand side are the eigenvalues of the diagonal 
twist matrix $\zt:=(\zt_1,\ldots, \zt_\ell)$ that is used to twist the the corresponding monodromy matrix $\mon(u)$. 

\paragraph{\textsf{Transfer-matrix eigenvalues}.}\hspace{-0.2cm}In the setting, where the local $L(u)$-operator is polynomial, the eigenvalues of the corresponding transfer matrix read as  
\bea
\la{eigenABA}
\mT(u)=\sum_{\a=1}^{\ell}\raisemath{-1pt}{\zt_\a}\DP_{\a}(u)
\frac{\Qa_{\a-1}(u-\hb)}{\Qa_{\a-1}(u)}\frac{\Qa_{\a}(u+\hb)}{\Qa_{\a}(u)}\, ,
\eea
The condition of polynomiality of $\mT(u)$ requires the residues of   (\ref{eigenABA}) to vanish, which results into the fulfilment of equations (\ref{Bethe_concise}) for 
Bethe roots $u_j^{(\a)}$.
Note that in the algebraic Bethe ansatz framework $\mT(u)$ comes out as the sum 
\bea
\la{tiL}
\mT(u)=\sum_{\a=1}^{\ell}\qeg_\a(u)\, ,
\eea
of the following $\ell$ terms 
\bea
\la{mcAB}
\qeg_{\a}(u)=\raisemath{-1pt}{\zt_\a}\DP_{\a}(u)
\frac{\Qa_{\a-1}(u-\hb)}{\Qa_{\a-1}(u)}\frac{\Qa_{\a}(u+\hb)}{\Qa_{\a}(u)}\, .
\eea
The quantities $\qeg_{\a}(u)$  can be interpreted as eigenvalues of the twisted monodromy matrix $\Mt=\zt\mon(u)$, see e.g. \cite{Arutyunov2026} and references therein, and they are 
commonly referred to as {\it quantum eigenvalues}. We assume that the twist matrix $\zt$ has simple eigenvalues.


\smallskip

For rank $\ell>2$, in addition to the transfer matrix $\mT_1(u):=\mT(u)$,  there are higher transfer matrices $\mT_{\xx}(u)$ with $\xx=2,\ldots, \ell$. Together with 
$\mT_1(u)$, they form a family of {\it fundamental quantum characters} constructed as traces of the monodromies over the fundamental representations  of $\gl_\ell$.
Explicitly, they are built via quantum eigenvalues by using the tableau sum formula \cite{Kuniba:2011}
\bea
\la{Miura}
\mT_\xx(u)=\sum_{1\leq \a_1< \ldots <\a_\xx\leq \ell}\qeg_{\a_1}(u)\cdots \qeg_{\a_\xx}(u+(\xx-1)\hb)\, ,
\eea
which is a transfer-matrix analogue of  the $\xx$-th elementary symmetric function $\mathsf{e}_\xx$ expressed via eigenvalues of a group element $g\in {\rm GL}_{\ell}$.
The quantum character $\mT_{\ell}(u)$ is the quantum determinant and, when regarded as an operator, generates the center of the $\gl_{\ell}$ Yangian. Upon substituting 
the quantum eigenvalues (\ref{mcAB}) in (\ref{Miura}), we obtain 
\bea
\la{BA_inspired}
\begin{aligned}
\mT_{\xx}(u)&=\sum_{1\leq \a_1< \ldots <\a_{\xx}\leq \ell}\zt_{\a_1}\ldots \zt_{\a_\xx}\DP_{\a_1}(u)\ldots \DP_{\a_\xx}(u+(\xx-1)\hb)\\
&\times \frac{Q_{\a_1-1}(u-\hb)}{Q_{\a_1-1}(u)}\frac{Q_{\a_1}(u+\hb)}{Q_{\a_1}(u)}\ldots \frac{Q_{\a_\xx-1}(u+(\xx-2)\hb)}{Q_{\a_\xx-1}(u+(\xx-1)\hb)}
\frac{Q_{\a_\xx}(u+\xx \hb)}{Q_{\a_\xx}(u+(\xx-1)\hb)}\, ,
\end{aligned}
\eea
which is valid for $\xx=1,\ldots, \ell$ and for $\xx=1$ yields the expression (\ref{eigenABA}). 
The operators corresponding to the eigenvalues (\ref{mcAB}) pairwise commute for different values of the 
spectral parameter. This implies the commutativity of the fundamental quantum characters. Their common eigenvalues are precisely given by (\ref{BA_inspired}).

\paragraph{\textsf{$\BB$-operator.}}\hspace{-0.2cm}An important ingredient of the new construction of the spin chain spectrum is the $\BB(u)$-operator \cite{Gromov:2016itr} that generalizes the Sklyanin construction 
for the $\gl_3$ spin chain for arbitrary rank $\ell$. It is given by
\bea
\la{B_atern_norm}
&&\BB(u)=c_{\ell}\,
\rv_{\a_1}\ldots \rv_{\a_{\ell-1}}\rv_{\a_{\ell}} \eps_{ \b_1\ldots \b_{\ell-1}\a_{\ell} }\\
\nonumber
&&\hspace{2cm}\times \, \Mt_{\a_1^1}^{\, \a_1}(u)\Mt_{\a_2^1\a_2^2}^{\a_1^1\, \a_2}(u)\Mt_{\a_3^1\a_3^2\a_3^3}^{\a_2^1\a_2^2\, \a_3}(u)\, \ldots \, \Mt_{\, \b_1\b_2
~\ldots~\b_{\ell-2}\b_{\ell-1}}^{\a_{\ell-2}^1 ~\ldots~  \a_{\ell-2}^{\ell-2}\, \a_{\ell-1}}(u)\, ,
\eea
where $\Mt_{\a_1\ldots \a_\xx}^{\beta_1\ldots \,\beta_\xx}(u)$ are \index{quantum minors}{\it quantum minors} constructed  from the entries of the twisted monodromy matrix $\Mt(u)$,
which satisfies the Yang-Baxter relations with the rational Yang $R$-matrix. Explicitly,
\bea
\begin{aligned}
\Mt_{\a_1\ldots \a_\xx}^{\beta_1\ldots  \, \beta_\xx}(u)
&=\sum_{\tau \in \Sym_\xx} \sgn\tau\, \Mt_{\a_1 \b_{\tau(1)}}(u)\ldots \Mt_{\a_k\b_{\tau(\xx)}}(u+(\xx-1)\hb)\, ,
\end{aligned}
\eea
where $ \Sym_\xx$ is a symmetric group on $\xx$ elements.
In (\ref{B_atern_norm}), the variables $\rv_\a$, $\a=1,\ldots, \ell$, are the parameters governing the normalization of the Baker-Akhieser function in classical theory, while 
$c_{\ell}$ is an appropriate normalization constant for $\BB(u)$. 
In classical theory, the zeros of the polynomial $\BB(u)$ coincide with the poles of the Baker-Akhieser function. In quantum theory, the operators $\BB(u)$ for different values of the 
spectral parameter commute, and their joint spectrum  is naturally labeled by  Gelfand-Tsetlin patterns \cite{Ryan:2020rfq}. For a twist $\zt$ with simple 
spectrum, $\BB(u)$ is a polynomial of degree $\scalebox{1.1}{$\mathpzc{n}$}\hspace{0.1mm}=N\ell(\ell-1)/2$. This degree coincides with the number of physical degrees of freedom of 
the $\gl_{\ell}$ spin chain of length $N$. 

The most important point about the $\BB(u)$ operator is that it generates the states of the spin chain 
\bea
\la{B_q_norm_opnew}
|v\rangle \propto \BB(v_1)\BB(v_2)\ldots \BB(v_{\mb_1})|\Omega\rangle\, ,
\eea
where $|\Omega\rangle$ denotes the vacuum eigenstate of $\mT(u)$ and $v_{k}$,  $k=1,\ldots, \mb_1$ are the Bethe roots of the first level.
Unlike the standard algebraic Bethe ansatz, where the construction of transfer-matrix eigenstates requires $\BB$-operators from higher (nested) levels,
the states (\ref{B_q_norm_opnew}) are generated solely by the unique operator $\BB(u)$. As a result, the auxiliary roots of the nested Bethe ansatz 
remain invisible in this approach, suggesting that the principal roots  $v_k$ should satisfy a closed system of equations.
Deriving this system for arbitrary rank $\ell$ is a main goal of the present paper. 

A straightforward way to derive such a system is to follow the logic of the algebraic Bethe ansatz. One applies the transfer matrix $\mT(u)$ to the state $|v\rangle$ and
commutes $\mT(u)$ through all $\BB$-operators until it reaches the vacuum state, which is an eigenstate of $\mT(u)$. In the process, one expects to obtain a term proportional to $|v\rangle$
with coefficient given by the transfer-matrix eigenvalue, together with a set of so-called unwanted terms. Requiring the unwanted terms to vanish then yields the desired equations. 
Unfortunately, the commutation relations between $\mT(u)$ and the $\BB$-operator are rather involved due to the intricate structure  of $\BB(u)$, see (\ref{B_atern_norm}).
Even in the simplest nontrivial case of the $\gl_3$ spin chain, the analytic proof  of the Bethe-vector construction  (\ref{B_q_norm_opnew})
requires considerable effort \cite{Liashyk:2018qfc,Liashyk:2019sym}. Therefore, to derive a closed system of equations for the roots $v_k$ 
of the on-shell Bethe vectors  (\ref{B_q_norm_opnew}),
we adopt a different approach based on the quantum spectral curve.

\paragraph{\textsf{Quantum spectral curve}.}\hspace{-0.2cm}We therefore turn to the quantum spectral curve construction. 
In the SoV framework developed in \cite{Maillet:2019ypa} (see also  \cite{Arutyunov2026}), the quantum spectral curve 
takes the form of the functional equation (Baxter's $\mT Q$-relation)
satisfied by the first Baxter polynomial $Q_1(u)$, namely, 
 \bea
\la{main_qsc_sep_gen}
\sum_{\xx=0}^{\ell}(-1)^{\ell-\xx}\,  \zt_1^\xx\,  \alg{a}_\xx(u)\mT_{\ell-\xx}(u+\xx\hb)Q_1(u+\xx\hb)=0\, ,
\eea
where $\alg{a}_0(u)=1$, $\mT_0(u)=1$, and for $\xx\geq 1$, the coefficients $\alg{a}_\xx(u)$ are given by
\bea
\la{new_ak}
\alg{a}_\xx(u)=\prod_{m=0}^{\xx-1}\DP_{1}(u+m\hb)\, ,~~~~\xx=1,\ldots, \ell\, ,
\eea
and $\alg{a}_0(u)=1$. In general, the quantum spectral curve equation has $\ell$ fundamental solutions $\gQ_\a$, $\a=1,\ldots, \ell$. However, only one of them 
coincides exactly with the first Baxter polynomial $\gQ_1=Q_1$. Other $\gQ_\a$ can be used to build higher Baxter's polynomials $Q_\a$, $\a>1$, but we do not need 
these relations here.

The quantum spectral curve equation (\ref{main_qsc_sep_gen}) is valid for the $\gl_{\ell}$ spin chain in {\it any} representation. 
For the spin chain in the vector representations, the coefficients $\alg{a}_\xx$, $\xx\geq 1$, are given by (\ref{new_ak}),
where $\DP_1(u)$ is defined in (\ref{DP}).

As shown in \cite{Arutyunov2026}, the quantum spectral curve equation (\ref{main_qsc_sep_gen}) admits multiple derivations. 
In the following sections, we derive it directly from (\ref{BA_inspired}).

\paragraph{\textsf{Lagrange interpolation for transfer matrices}.}\hspace{-0.2cm}Our construction is completed by the  Lagrange interpolation formula.
Recall that $\mT_\xx(u)$ is a polynomial of degree $\xx N$ in the spectral parameter $u$, with kinematic
zeros at $u=u_j-n\eta$,  for $n=1,\ldots, \xx-1$, and asymptotic behavior 
as $u\to \infty$:
\bea
\mT_{\xx}(u)\sim \chi^{(\ell)}_{\raisebox{-0.4ex}{$\scriptstyle \xx$}}(\zt) u^{\xx N}
\, .
\eea
Here $\chi^{(\ell)}_{\raisebox{-0.4ex}{$\scriptstyle \xx$}}(\zt)$ denotes the character of the $\xx$-th fundamental representation of $\gl_\ell$. 
Equivalently,  it is the $\xx$-th
elementary symmetric function of $\zt_1,\ldots,\zt_{\ell}$, namely,
\bea
\chi^{(\ell)}_{\raisebox{-0.4ex}{$\scriptstyle \xx$}}(\zt)=\sum_{1\leq \a_1<\ldots <\a_{\xx}\leq \ell} \zt_{\a_1}\ldots \zt_{\a_{\xx}}\, .
\eea
In the untwisted case, $\zt_1=\zt_2=\ldots =\zt_{\ell}=1$, the character reduces to the dimension of the
representation, $\chi^{(\ell)}_{\raisebox{-0.4ex}{$\scriptstyle \xx$}}(1)=\binom{\ell}{\xx}$.  

It turns out that this data is exactly sufficient to uniquely reconstruct the polynomial $\mT_\xx(u)$,  
provided the values $\mT_\xx(u_j)$ for all $j=1,\ldots, N$  are known.
To this end, we introduce an auxiliary function $g_\xx(u)$ that encodes kinematic zeros of the transfer matrix
\bea
\la{def_g}
g_\xx(u)=\prod_{j=1}^N\prod_{m=1}^{\xx-1}(u-u_j+m\eta)\, 
\eea
and $g_1(u)=1$.
The Lagrange interpolation formula for $\mT_\xx(u)$ yields 
\bea
\la{LIT}
\mT_\xx(u)=g_\xx(u)\left[\chi^{(\ell)}_{\raisebox{-0.4ex}{$\scriptstyle \xx$}}(\zt)\prod_{j=1}^N (u-u_j)+\sum_{j=1}^{N}\frac{\mT_\xx(u_j)}{g_\xx(u_j)}\prod_{s\neq j}^N\frac{u-u_s}{u_j-u_s}\right]\, .
\eea
In particular, 
\bea
\la{LIT1}
\mT_1(u)=\chi^{(\ell)}_{\raisemath{-0.5ex}{1}}(\zt)\prod_{j=1}^N(u-u_j)+\sum_{j=1}^N \mT_1(u_j) \prod_{s\neq j}^N \frac{u-u_s}{u_j-u_s}\, ,
\eea
where $\chi_{\raisemath{-0.5ex}{1}}^{(\ell)}(\zt)={\rm Tr}\zt$. Apparently, the only rank-dependent quantities in these formulae are the character $\chi^{(\ell)}_k(\zt)$.\footnote{Unless stated 
otherwise, we suppress the rank label and write $\chi_{\raisebox{-0.4ex}{$\scriptstyle \xx$}}$ instead of $\chi_{\raisebox{-0.4ex}{$\scriptstyle \xx$}}^{(\ell)}$.  Whenever several ranks appear simultaneously, the superscript is restored. }
Further discussion of these formulae can be found in \cite{Maillet:2019ypa,Arutyunov2026}.

In the untwisted case, the transfer matrix is $\gl_{\ell}$-invariant, and the spectrum of $\mT_{1}(u_j)$, considered as an operator, is degenerate, splitting into families 
of coincident eigenvalues. The number of such eigenvalues equals the dimension of an irreducible $\gl_{\ell}$-module that appears in the decomposition of $({\mathbb C}^{\ell})^{\otimes N}$
into irreducible components.

\section{Non-nested Bethe Equations for the $\gl_3$ Spin Chain}
\paragraph{\textsf{Transfer-matrix eigenvalues and fusion relations.}}\hspace{-0.2cm}We begin with the simplest nontrivial example of the $\gl_3$ spin chain. Applying relations  (\ref{BA_inspired}) for this chain,   we obtain
\bea
\la{T1T2}
\begin{aligned}
\mT_1(u)&=\zt_1\DP_1(u)\frac{Q_1(u+\hb)}{Q_1(u)}+\zt_2\DP_2(u)\frac{Q_1(u-\hb)}{Q_1(u)}\frac{Q_2(u+\hb)}{Q_2(u)}+\zt_3\DP_3(u)\frac{Q_2(u-\hb)}{Q_2(u)}\, ,\\
\mT_2(u)&=\zt_1\zt_2\DP_1(u)\DP_2(u+\hb)\frac{Q_2(u+2\hb)}{Q_2(u+\hb)}\\
&+\zt_1\zt_3\DP_1(u)\DP_3(u+\hb)\frac{Q_1(u+\hb)}{Q_1(u)}\frac{Q_2(u)}{Q_2(u+\hb)}
+\zt_2\zt_3\DP_2(u)\DP_3(u+\hb)\frac{Q_1(u-\hb)}{Q_1(u)}\, , \\
\mT_3(u)&=\zt_1\zt_2 \zt_3\DP_1(u)\DP_2(u+\hb)\DP_3(u+2\hb)\, ,
\end{aligned}
\eea
where $\DP_\a(u)$ for $\a=1,2,3$, are vacuum polynomials (\ref{DP}). The transfer matrix  $\mT_3(u)$ does not involve Baxter's polynomials
and represents the value of the quantum determinant.  

We can now see that the equations in (\ref{T1T2}) imply the following identity
\bea
\la{func_ansatz_solv_3}
\begin{aligned}
&\zt_1^3\alg{a}_3(u)Q_1(u+3\hb)- \zt_1^2\alg{a}_2(u)\mT_1(u+2\hb)Q_1(u+2\hb)
\\&\hspace{3cm}~
+ \zt_1\alg{a}_1(u)\mT_{2}(u+\hb)Q_1(u+\hb)-\mT_{3}(u)Q_1(u)=0\, ,
\end{aligned}
\eea
where the coefficients $\alg{a}_\xx(u)$ are given in (\ref{new_ak}). Equation (\ref{func_ansatz_solv_3}) coincides with the quantum spectral curve 
(\ref{main_qsc_sep_gen}) for $\ell=3$. As we have pointed out above, 
formulae (\ref{T1T2}) and (\ref{func_ansatz_solv_3}) hold for the $\gl_3$ spin chain in {\it any} representation. 
One distinctive feature of the vector representation is the existence of specific fusion relations among transfer matrices. 
In other words, for the spin chain in the vector representation, all spectral information is encoded in a single polynomial
$Q_1(u)$, which is generally not the case for other representations. 

To derive the fusion relations in the present context, we consider the vector representation at each site of the $\gl_3$ chain. In this case, the vacuum polynomials are given by 
(\ref{DPv}).  From (\ref{T1T2}) we obtain
\bea
\la{T_on_inhom}
\begin{aligned}
&\mT_1(u_j)=\zt_1 \DP_1(u_j)\frac{Q_1(u_j+\eta)}{Q_1(u_j)}\, , \\
&\mT_1(u_j+\eta)=\zt_2\DP_2(u_j+\eta)\frac{Q_1(u_j)}{Q_1(u_j+\eta)}\frac{Q_2(u_j+2\hb)}{Q_2(u_j+\eta)}+\zt_3\DP_3(u_j+\eta)\frac{Q_2(u_j)}{Q_2(u_j+\eta)}\, , \\
&\mT_2(u_j)=\zt_1\zt_2\DP_1(u_j)\DP_2(u_j+\hb)\frac{Q_2(u_j+2\hb)}{Q_2(u_j+\hb)}\\
&\hspace{2cm}+\zt_1\zt_3\DP_1(u_j)\DP_3(u_j+\hb)\frac{Q_1(u_j+\hb)}{Q_1(u_j)}\frac{Q_2(u_j)}{Q_2(u_j+\hb)}\, ,
\end{aligned}
\eea
where $j=1,\ldots, N$ and we have used the fact that $\DP_2(u_j)=\DP_3(u_j)=0$. Using $\DP_1(u_j+\eta)=0$, we also obtain 
\bea
\mT_2(u_j)=\zt_2\zt_3\DP_2(u_j+\eta)\DP_3(u_j+2\hb)\frac{Q_1(u_j)}{Q_1(u_j+\eta)}\, .
\eea
From these formulae we immediately deduce that 
\bea
\begin{aligned}
&\mT_1(u_j)\mT_1(u_j+\eta)=\mT_2(u_j)\, ,\\
&\mT_1(u_j)\mT_2(u_j+\eta)=\mT_3(u_j)\, , \\
&\mT_3(u_j+\eta)=0\, ,
\end{aligned}
\eea
where the last relation again follows from $\DP_1(u_j+\eta)=0$. These are the fusion relations for the $\gl_3$ transfer matrices evaluated on inhomogeneitis. 
More generally, for the $\gl_\ell$ spin chain in the vector representation they are given by the following recurrent relations
\bea
\la{spec_sys_sep}
\begin{aligned}
&\mT_1(u_j)\mT_{\xx}(u_j+\hb)=\mT_{\xx+1}(u_j)\, ,~~~~~\xx=1,\ldots \ell-1, \\
&\mT_{\ell}(u_j+\hb)=0\, .
\end{aligned}
\eea
In fact, these relations can be derived in a variety of different ways, and can be effectively used for finding the transfer matrix spectrum 
providing an alternative to the standard Bethe ansatz construction, see e.g. \cite
{Maillet:2019ypa,Arutyunov2026}. Here we demonstrated that these relations is a simple consequence of the representation of quantum characters via quantum eigenvalues.

\paragraph{\textsf{Derivation of equations for principal Bethe roots}.}\hspace{-0.2cm}Now we turn to the main result: the derivation of a closed system of equations for the principal Bethe roots $v_k$. Using the fact that $v_k$, $k=1,\ldots , \mb$, are roots of 
$Q_1(v)$, i.e., $Q_1(v_k)=0$, we evaluate  equation (\ref{func_ansatz_solv_3}) at $u=v_k-\eta$, obtaining 
\bea
\la{func_ansatz_solv_exp}
\begin{aligned}
&\zt_1^3\alg{a}_3(v_k-\eta)Q_1(v_k+2\hb)- \zt_1^2\alg{a}_2(v_k-\eta)\mT_1(v_k+\hb)Q_1(v_k+\hb)
\\&\hspace{6cm}~
-\mT_{3}(v_k-\eta)Q_1(v_k-\eta)=0\, .
\end{aligned}
\eea
Note that all the quantities appearing in this equation are finite. Substituting the expression for $\mT_3(u)$ from (\ref{T1T2}) and using (\ref{new_ak}) together with the vacuum polynomials (\ref{DPv}),
we rewrite the equation as
\bea
\la{alm_B}
Q_1(v_k+2\hb)-\frac{\mT_1(v_k+\hb)}{\zt_1\DP_1(v_k+\hb)}Q_1(v_k+\hb)=\frac{\chi_3(\zt)}{\zt_1^3}\prod_{j=1}^N\frac{v_k-u_j+\hb}{v_k-u_j-\hb}Q_1(v_k-\hb)\, ,
\eea
where $\chi_3(\zt)=\zt_1\zt_2\zt_3$. The only remaining unknown quantity in this expression is the transfer matrix $\mT_1(v_k+\hb)$. To eliminate it, 
we first apply  the Lagrange interpolation formula  (\ref{LIT1}) to obtain
\bea
\la{LIT1ev}
\frac{\mT_1(v_k+\hb)}{\zt_1\DP_1(v_k+\hb)}=\frac{\chi_1(\zt)}{\zt_1}\prod_{j=1}^N\frac{v_k-u_j+\eta}{v_k-u_j}+\sum_{j=1}^N \frac{\mT_1(u_j)}{\zt_1\DP_1(v_k+\hb)} \prod_{s\neq j}^N \frac{v_k-u_s+\eta}{u_j-u_s}\, ,
\eea
where $\chi_1(\zt)=\zt_1+\zt_2+\zt_3$.
We then substitute the expression for  $\mT_1(u_j)$ given in the first line of (\ref{T_on_inhom}), namely, 
\bea
\la{imp_vspom}
\mT_1(u_j)=\zt_1 \DP_1(u_j)\frac{Q_1(u_j+\eta)}{Q_1(u_j)}\, .
\eea
 Rearranging the terms, we obtain  
\bea
\la{univ}
\frac{\mT_1(v_k+\hb)}{\zt_1\DP_1(v_k+\hb)}
=\prod_{j=1}^N \frac{v_k-u_j+\hb}{v_k-u_j}
 \bigg\{\frac{\chi_1(\zt)}{\zt_1}+\sum_{l=1}^N \DP_1(u_l)h(v_k,u_l)\frac{Q_1(u_l+\hb)}{Q_1(u_l)}\bigg\}\, ,
\eea
where we have introduced the auxiliary functions $h(u,u_l)$ \cite{Arutyunov2026},
\bea
\la{huu_l}
h(u,u_l)=\frac{1}{u-u_l+\eta}\prod_{m\neq l}^N \frac{1}{u_l-u_m}\, .
\eea
Substituting (\ref{univ}) into (\ref{alm_B}), we obtain 
\bea
\la{alm_B_1}
\begin{aligned}
\frac{Q_1(v_k+2\hb)}{ Q_1(v_k+\hb)}\prod_{j=1}^N \frac{v_k-u_j}{v_k-u_j+\hb} &-\frac{\chi_3(\zt)}{\zt_1^3}\frac{Q_1(v_k-\hb)}{ Q_1(v_k+\hb)}\prod_{j=1}^N\frac{v_k-u_j}{v_k-u_j-\hb}
\\
&=\frac{\chi_1(\zt)}{\zt_1}+\sum_{l=1}^N \DP_1(u_l)h(v_k,u_l)\frac{Q_1(u_l+\hb)}{Q_1(u_l)}\, .
\end{aligned}
\eea
where $k=1,\ldots, \mb_1$. Since only the first Baxter polynomial, whose roots are $v_k$, is involved,
this constitutes a closed system of equations for the principal Bethe roots. 
Its admissible solutions\footnote{A solution of (\ref{alm_B_1}) is admissible if it does not contain coincident roots.} 
determine the transfer-matrix spectrum of the $\gl_3$ spin chain in the vector representation
via (\ref{B_q_norm_opnew}). 
\medskip

\paragraph{\textsf{Interpolation and resummation.}}\hspace{-0.2cm}Once the system has been  solved, the transfer-matrix eigenvalues can be computed from (\ref{imp_vspom})
together with the Lagrange interpolation formula (\ref{LIT1}) for $\mT_1(u)$, namely,
\bea
\la{see_T!}
\hspace{-0.3cm}\mT_1(u)=\chi_1(\zt)\prod_{j=1}^N(u-u_j)+\zt_1\sum_{j=1}^N \prod_{l=1}^N(u_j-u_l-\hb)\prod_{i\neq j}\frac{u-u_i}{u_j-u_i}\prod_{n=1}^{\mb_1}\frac{u_j-v_n+\hb}{u_j-v_n}\, .
\eea

Both equations (\ref{alm_B_1})  and the representation (\ref{see_T!}) suffer from a certain drawback: they do not possess a well-defined 
homogeneous limit $u_j\to 0$ for all $j=1,\ldots, N$. Indeed, the summation term appearing
on the right hand side of (\ref{alm_B_1}) can be written explicitly as 
\bea
&&\sum_{l=1}^N \DP_1(u_l)h(v_k,u_l)\frac{Q_1(u_l+\hb)}{Q_1(u_l)}=\\
\nonumber
&&
\hspace{3cm}-\sum_{l=1}^N\frac{\hb}{v_k-u_l+\eta}\prod_{j\neq l}^N \frac{u_l-u_j-\hb}{u_l-u_j}\prod_{n=1}^{\mb_1} \frac{u_l-v_n+\hb}{u_l-v_n}\, ,
\eea
which becomes singular when the inhomogeneities coincide. To obtain a representation that remains regular as $u_j\to 0$,
 we perform the following resummation.
 
Introduce the following function 
\bea
F(z)=\frac{1}{v_k-z+\hb}\prod_{j=1}^N\frac{z-u_j-\hb}{z-u_j}\prod_{n=1}^{\mb_1}\frac{z-v_n+\hb}{z-v_n}\, .
\eea
This function has poles at $z=u_j$ for any $j=1,\ldots, N$; at $z=v_k+\hb$ for any $k=1,\ldots, \mb_1$; and at $z=v_n$. There is also a pole at infinity. 
Consider the Cauchy integral of $F(z)$ over a contour that encircles all the poles in the finite part of the complex plane
\bea
\la{around_poles}
\begin{aligned}
&\frac{1}{2\pi \i}\oint_{\rm poles}F(z)dz=-\sum_{l=1}^N\frac{\hb}{v_k-u_l+\eta}\prod_{j\neq l}^N \frac{u_l-u_j-\hb}{u_l-u_j}\prod_{n=1}^{\mb_1} \frac{u_l-v_n+\hb}{u_l-v_n}\\
&-\prod_{j=1}^N\frac{v_k-u_j}{v_k-u_j+\hb}\prod_{n=1}^{\mb_1}\frac{v_k-v_n+2\hb}{v_k-v_n+\hb}
+\sum_{n=1}^{\mb_1}\frac{\hb}{v_k-v_n+\hb}\prod_{j=1}^N\frac{v_n-u_j-\hb}{v_n-u_j}\prod_{s\neq n}^{\mb_1}\frac{v_n-v_s+\hb}{v_n-v_s}\, ,
\end{aligned}
\eea
where the right-hand side is the sum of the corresponding residues.
Alternatively, the same integral can be evaluated by computing the residue at infinity, which produces
\bea
\la{int_inf}
\frac{1}{2\pi \i}\oint_{\rm \infty}F(z)dz=-1\, .
\eea
Equating (\ref{around_poles}) and (\ref{int_inf}) yields the following identity 
\bea
\la{Id_imp}
\begin{aligned}
\sum_{l=1}^N \DP_1(u_l)h(v_k,u_l)\frac{Q_1(u_l+\hb)}{Q_1(u_l)}&=-1+\frac{Q_1(v_k+2\hb)}{ Q_1(v_k+\hb)}\prod_{j=1}^N \frac{v_k-u_j}{v_k-u_j+\hb}\\
&-\sum_{n=1}^{\mb_1}\frac{\hb}{v_k-v_n+\hb}\prod_{j=1}^N\frac{v_n-u_j-\hb}{v_n-u_j}\prod_{s\neq n}^{\mb_1}\frac{v_n-v_s+\hb}{v_n-v_s}\, .
\end{aligned}
\eea
The identity (\ref{Id_imp}) can be verified numerically for generic choices of the variables
$\{v_k\}$ and the inhomogeneity parameters $\{u_j\}$. Substituting this result into
(\ref{alm_B_1}), we obtain the alternative representation
\bea
\la{alm_B_2}
\begin{aligned}
\frac{\zt_2+\zt_3}{\zt_1}+\frac{\zt_2\zt_3}{\zt_1^2}&\prod_{j=1}^N\frac{v_k-u_j}{v_k-u_j-\hb}\prod_{n=1}^{\mb_1}\frac{v_k-v_n-\hb}{v_k-v_n+\hb}
\\
=&\, \sum_{n=1}^{\mb_1}\frac{\hb}{v_k-v_n+\hb}\prod_{j=1}^N\frac{v_n-u_j-\hb}{v_n-u_j}\prod_{s\neq n}^{\mb_1}\frac{v_n-v_s+\hb}{v_n-v_s}\, .
\end{aligned}
\eea
which is algebraically equivalent to the original expression. Unlike (\ref{alm_B_1}),
however, the form (\ref{alm_B_2}) is manifestly regular in the homogeneous limit
$u_j\to0$. The apparent singularities arising from the Lagrange interpolation
representation are reorganized into combinations that remain finite in this
limit. Consequently, (\ref{alm_B_2}) provides a more convenient starting point for both
analytical and numerical studies of the model. Numerical evidence that the non-nested equations (\ref{alm_B_2}) generate the same 
set of solutions for the principle Bethe roots as the standard nested Bethe equations is presented in appendix A. 

A similar resummation can be performed for the transfer-matrix eigenvalue
formula (3.14). To this end, we introduce the auxiliary meromorphic function
\bea
G(z)=\frac{1}{u-z}\prod_{j=1}^N\frac{u-u_j}{z-u_j}\prod_{j=1}^N(z-u_l-\hb)\prod_{n=1}^{\mb_1}\frac{z-v_n+\hb}{z-v_n}\, .
\eea
The function possesses simple
poles at $z=u_j$, $j=1,\ldots,N$, at the Bethe roots $z=v_n$,
$n=1,\ldots,m_1$, and at the additional point $z=u$. Furthermore, since the
degree of the denominator exceeds that of the numerator by one, $G(z)$ also
has a pole at infinity.

Evaluating the contour integral of $G(z)$ both as a sum of residues at finite poles and as the residue at infinity, and then equating the results, yields the following identity
\bea
\nonumber
&&\hspace{-0.8cm}\sum_{j=1}^N \prod_{l=1}^N(u_j-u_l-\hb)\prod_{i\neq j}\frac{u-u_i}{u_j-u_i}\prod_{n=1}^{\mb_1}\frac{u_j-v_n+\hb}{u_j-v_n}
=\\
\la{PFD}
&&-\prod_{j=1}^N(u-u_j)+\prod_{j=1}^N(u-u_j-\hb)\prod_{n=1}^{\mb_1}\frac{u-v_n+\hb}{u-v_n}\\
\nonumber
&&\hspace{2cm}-\sum_{k=1}^{\mb_1}\frac{\hb}{u-v_k}\prod_{j=1}^N \frac{u-u_j}{v_k-u_j}\prod_{j=1}^N(v_k-u_j-\hb)\prod_{n\neq k}^{\mb_1} \frac{v_k-v_n+\hb}{v_k-v_n}\, .
\eea
which relates the contributions from the finite poles to the residue at
infinity. Again, this identity can be verified numerically for generic choices of the Bethe roots
and the inhomogeneity parameters.  

Finally, substituting (\ref{PFD})  into the interpolation formula (\ref{see_T!}) yields
at
\bea
\la{T_gl_2_adm_hom}
\begin{aligned}
\mT_1(u)=(\zt_2+\zt_3)\prod_{j=1}^N(u-u_j)+\zt_1&\prod_{j=1}^N(u-u_j-\hb)\prod_{n=1}^{\mb_1}\frac{u-v_n+\hb}{u-v_n}\\
-\zt_1\sum_{k=1}^{\mb_1}\frac{\hb}{u-v_k}&\prod_{j=1}^N(v_k-u_j-\hb)\prod_{j=1}^N \frac{u-u_j}{v_k-u_j}\prod_{n\neq k}^{\mb_1} \frac{v_k-v_n+\hb}{v_k-v_n}\, ,
\end{aligned}
\eea 
which provides an alternative representation of the transfer-matrix
eigenvalue. Unlike (\ref{see_T!}), the expression (\ref{T_gl_2_adm_hom}) admits a smooth
homogeneous limit. It can be written more compactly as 
\bea
\la{newT1}
\mT_1(u)=\zt_1\DP_1(u)\frac{Q_1(u+\hb)}{Q_1(u)}+\DP_1(u+\hb)\mathcal{A}(u-\hb)\, ,
\eea
where we have introduced 
\bea
\la{Anew}
\mathcal{A}(u)=\bar{\chi}_1(\zt)-\zt_1\sum_{k=1}^{\mb_1}\frac{\hb  \mathcal{C}_k}{u-v_k+\hb}\, , ~~~\mathcal{C}_k=\frac{\DP_1(v_k)}{\DP_1(v_k+\hb)}\prod_{n\neq k}^{\mb_1}\frac{v_k-v_n+\hb}{v_k-v_n}\, ,
\eea
where $\bar{\chi}_1(\zt)=\chi_1(\zt)-\zt_1$. For the present case of $\gl_3$, we have $\bar{\chi}_1(\zt)=\zt_2+\zt_3$.
\smallskip

Analogously, we reconstruct the transfer matrix $\mT_2(u)$. Using the Lagrange interpolation formula (\ref{LIT})  for $\mT_2(u)$, we express the values $\mT_2(u_j)$ through the fusion relation 
\bea
\la{RV}
\mT_2(u_j)=\mT_1(u_j)\mT_1(u_j+\hb),
\eea
where formula (\ref{T_gl_2_adm_hom})   is used for $\mT_1(u)$. After algebraic rearrangement similar to that performed for $\mT_1(u)$,
we obtain 
\bea
\nonumber
&&\mT_2(u)=\DP_1(u+2\hb)\bigg[ \zt_1\DP_1(u)\frac{Q_1(u+\hb)}{Q_1(u)}\mathcal{A}(u)\\
&&\hspace{5cm}+\DP_1(u+\hb)\bigg(\bar{\chi}_2 -\zt_1\sum_{k=1}^{\mb_1}\frac{\hb \mathcal{C}_k \mathcal{A}(v_k)}{u-v_k}\bigg)\bigg]\, , 
\la{T2_res}
\eea
where $\bar{\chi}_2(\zt)=\chi_2(\zt)-(\chi_1-\zt_1)\zt_1$. For $\gl_3$ we have $\bar{\chi}_2(\zt)=\zt_2\zt_3$.
Formula (\ref{T2_res}) exhibits the correct asymptotic behavior  $\mT_2(u)\sim \chi_2(\zt)u^{2N}$ as $u\to \infty$. The apparent pole families, $u=v_k$ and $u=v_k-\hb$, cancel. 
Thus $\mT_2(u)$ is indeed polynomial.

\medskip

\paragraph{\textsf{Embedding  property}.} \hspace{-0.2cm}We now examine equation (\ref{alm_B_2}) in more detail. First, we rewrite it as 
\bea
\la{alm_B_mod}
\begin{aligned}
\frac{\zt_2+\zt_3}{\zt_1}+&\frac{\zt_2\zt_3}{\zt_1^2}\prod_{j=1}^N\frac{v_k-u_j}{v_k-u_j-\hb}\prod_{n=1}^{\mb_1}\frac{v_k-v_n-\hb}{v_k-v_n+\hb}
\\
=&\, -\sum_{n=1}^{\mb_1}\frac{\hb}{v_k-v_n+\hb}\prod_{s\neq n}^{\mb_1}\frac{v_n-v_s-\hb}{v_n-v_s} \prod_{j=1}^N\frac{v_n-u_j-\hb}{v_n-u_j}
\prod_{l=1}^{\mb_1}\frac{v_n-v_l+\hb}{v_n-v_l-\hb}\, .
\end{aligned}
\eea
It is convenient to introduce the  function
\bea
\la{fv}
f(v)=\prod_{j=1}^N\frac{v-u_j}{v-u_j-\hb}\prod_{n=1}^{\mb_1}\frac{v-v_n-\hb}{v-v_n+\hb}\, .
\eea
Setting $\gg_k:=f(v_k)$, 
equation (\ref{alm_B_mod}) can be rewritten as
\bea
\la{alm_B_mod1}
\begin{aligned}
\frac{\zt_2+\zt_3}{\zt_1}+&\frac{\zt_2\zt_3}{\zt_1^2}\gg_k= -\sum_{n=1}^{\mb_1}\frac{1}{\gg_n}\frac{\hb}{v_k-v_n+\hb}\prod_{s\neq n}^{\mb_1}\frac{v_n-v_s-\hb}{v_n-v_s}\, .
\end{aligned}
\eea
For generic pairwise distinct variables $v_k$, there exists an identity\footnote{This identity is well known in the theory of the rational Ruijsenaars-Schneider model, see, e.g., \cite{Arutyunov2026}.} 
\bea
\la{RSident}
\sum_{n=1}^{\mb_1}\frac{\hb}{v_k-v_n+\hb}\prod_{s\neq n}^{\mb_1}\frac{v_n-v_s-\hb}{v_n-v_s}=1\, .
\eea
This identity implies the existence of a special solution to (\ref{alm_B_mod}) for which all $\gg_k$ are equal, i.e., $\gg_k=\gg$ for $k=1,\ldots, \mb_1$.
In this case, equation (\ref{alm_B_mod1}) reduces to 
\bea
\la{quadr}
\frac{\zt_2\zt_3}{\zt_1^2}\gg^2+\frac{\zt_2+\zt_3}{\zt_1}\gg+1=0\, ,
\eea
which has two solutions
\bea
\la{sol_quadr}
\gg=-\frac{\zt_1}{\zt_2}\, , ~~~~\gg=-\frac{\zt_1}{\zt_3}\, .
\eea
These two solutions are essentially equivalent. Choosing the first one, the relation $\gg=f(v_k)$ becomes
\bea
\la{sl2_eq}
\frac{\zt_1}{\zt_2}\prod_{j=1}^N\frac{v_k-u_j-\hb}{v_k-u_j}=\prod_{n\neq k}^{\mb_1}\frac{v_k-v_n-\hb}{v_k-v_n+\hb}\, ,
\eea
which are exactly the Bethe equations for the twisted $\gl_2$ spin chain. Thus, the constant solutions  of (\ref{alm_B_mod1}) correspond to the reduction 
to the $\gl_2$ spin chain and manifest the embedding $\gl_2\subset \gl_3$.

We conclud our discussion of (\ref{alm_B_mod1}) by noting that these equations involve the Cauchy kernel $\hb/(v_k-v_n+\hb)$.
Applying its inverse, (\ref{alm_B_mod1}) can be written in the equivalent dual form 
\bea
\frac{\zt_2+\zt_3}{\zt_1}+\frac{1}{\gg_k}=\frac{\zt_2\zt_3}{\zt_1^2}\sum_{n=1}^{\mb_1}\gg_n\frac{\hb}{v_k-v_n-\hb} \prod_{s\neq n}^{\mb_1} \frac{v_n-v_s+\hb}{v_k-v_s}\, .
\eea
In section 5, we generalize this dual formulation to non-nested equations of arbitrary rank.

\paragraph{\textsf{The number of auxiliary roots and ${\rm U}(1)$ charges.}} \hspace{-0.2cm}Returning to the general equation (\ref{alm_B_mod}), one might think that information about nesting and the representation content of the transfer-matrix eigenvalues 
has disappeared. However, this is not the case. We now show how to recover the number $\mb_2$ of auxiliary Bethe roots  in the conventional nested Bethe ansatz 
corresponding to a given solution $(v_1,\ldots , v_{\mb_1})$ of  (\ref{alm_B_mod}). To this end, we equate the representations 
(\ref{newT1}) and  (\ref{T1T2})  for $\mT_1(u)$, obtaining
\bea
\mathcal{A}(u)=\zt_2\frac{Q_1(u)}{Q_1(u+\hb)}\frac{Q_2(u+2\hb)}{Q_2(u+\hb)}+\zt_3\frac{Q_2(u)}{Q_2(u+\hb)}\, .
\eea
Expanding the right-hand side in the limit $u\to \infty$, one finds 
\bea
\la{AA1}
\mathcal{A}(u)=\zt_1+\zt_3-\frac{\hb}{u}\big(\zt_2(\mb_1-\mb_2)+\zt_3\mb_2  \big)+\mathcal{O}(u^{-2})\, .
\eea 
This is to be compared with the expansion of (\ref{Anew}):
\bea
\la{AA2}
\mathcal{A}(u)=\zt_1+\zt_3-\zt_1\frac{\hb}{u}\sum_{k=1}^{\mb_1} \mathcal{C}_k+\mathcal{O}(u^{-2})\, .
\eea
Equating the coefficients of $u^{-1}$ in (\ref{AA1}) and (\ref{AA2}) yields
\bea
\la{AA3}
\begin{aligned}
\mb_2&=\frac{\zt_2}{\zt_2-\zt_3}\mb_1-\frac{\zt_1}{\zt_2-\zt_3}\sum_{k=1}^{\mb_1} \mathcal{C}_k\\
&=
\frac{\zt_2}{\zt_2-\zt_3}\mb_1-\frac{\zt_1}{\zt_2-\zt_3}\sum_{k=1}^{\mb_1}\frac{\DP_1(v_k)}{\DP_1(v_k+\hb)}\prod_{n\neq k}^{\mb_1}\frac{v_k-v_n+\hb}{v_k-v_n}\, .
\end{aligned}
\eea
Quite remarkably, numerical investigations indicate that, for solutions of (\ref{alm_B_2}),
the right-hand side of (\ref{AA3}) is always an integer. Analytically 
this would follow from the existence of a polynomial $Q_2(u)$ solving the second-level Baxter equation 
\bea
\zt_2 Q_1(u-\hb)Q_2(u+\hb)+\zt_3 Q_1(u) Q_2(u-\hb)=\mathcal{A}(u)Q_1(u) Q_2(u)\, .
\eea

Although equation (\ref{AA3}) becomes singular in the untwisted limit, the integers $\mb_1$ and $\mb_2$ 
determine the highest-weight labels 
$[M_1,M_2,M_3]$ of the corresponding $\gl_3$ multiplet in this limit, namely,
\bea
M_1=N-\mb_1\, , ~~~M_2=\mb_1-\mb_2\, , ~~~~M_3=\mb_2\, .
\eea

Now we turn to the derivation of equations similar to  (\ref{alm_B_2}) for the $\gl_4$ case.

\section{Non-nested Bethe Equations for the $\gl_4$ Spin Chain}
\paragraph{\textsf{Derivation of equations for principal Bethe roots}.}\hspace{-0.2cm}For the $\gl_4$ spin chain the quantum spectral curve equation (\ref{main_qsc_sep_gen}) reads as
\bea
\la{QSC4}
   && \zt_1^4\mathfrak{a}_4(u) Q_1(u+4\eta)-\zt_1^3\mathfrak{a}_3(u)\mT_1(u+3\eta)Q_1(u+3\eta)\\ \nonumber
&&\hspace{0.3cm}+\zt_1^2\mathfrak{a}_2(u)\mT_2(u+2\eta)Q_1(u+2\eta)-\zt_1\mathfrak{a}_1(u)\mT_3(u+\eta)Q_1(u+\eta)+\mT_4(u)Q_1(u)=0\, ,
\eea
where we specify the vacuum polynomials and hence the coefficients $\mathfrak{a}_\xx(u)$ to the vector representation of $\gl_4$.
To derive the equations for the pricipal Bethe roots $v_k$, we evaluate the quantum spectral curve (\ref{QSC4}) at 
$u=v_k-\hb$. Using $Q_1(v_k)=0$, the resulting relation can be written in the form
\begin{align}
    A_k-B_k+C_k+D_k=0\, ,
\end{align}
where
\bea
\la{ABCD}
\begin{aligned}
    A_k
    &:=
    \frac{Q_1(v_k+3\eta)}{Q_1(v_k+\eta)}\frac{\DP_1(v_k+\hb)}{\DP_1(v_k+3\hb)}\, ,
    \\
    B_k
    &:=
    \frac{\mT_1(v_k+2\eta)}
    {\zt_1\DP_1(v_k+2\eta)}\frac{Q_1(v_k+2\eta)}{Q_1(v_k+\eta)}
   \frac{\DP_1(v_k+\hb)}{\DP_1(v_k+3\hb)},
    \\
    C_k
    &:=
    \frac{\mT_2(v_k+\eta)}
    {\zt_1^2\DP_1(v_k+\eta)\DP_1(v_k+2\eta)}
  \frac{\DP_1(v_k+\hb)}{\DP_1(v_k+3\hb)},
    \\
    D_k
    &:=
    \frac{\mT_4(v_k-\eta)}
    {\zt_1^4\DP_1(v_k-\eta)\DP_1(v_k)\DP_1(v_k+\eta)\DP_1(v_k+2\eta)}\frac{Q_1(v_k-\eta)}{Q_1(v_k+\eta)}
       \frac{\DP_1(v_k+\hb)}{\DP_1(v_k+3\hb)}.
\end{aligned}
\eea
The common factor on the right-hand side of these expressions has been introduced for further convenience.
Since 
\bea
\mT_4(u)=\chi_4(\zt)\prod_{i=1}^4\DP_i(u)\, ,  
\eea
the coefficient $D_k$ is given by
\bea
\la{Dev}
 D_k
    =
    \frac{\chi_4(\zt)}{\delta_1^4} \frac{Q_1(v_k-\eta)}{Q_1(v_k+\eta)}\frac{\DP_1(v_k+\hb)}{\DP_1(v_k)}\, .
\eea
The formulae (\ref{newT1}) and (\ref{T2_res})  were derived solely from the interpolation formula (\ref{LIT}) and the fusion relation (\ref{RV}). 
Therefore, they depend on the rank $\ell$ of $\gl_{\ell}$ only through the characters. Replacing
the $\gl_3$ characters by those of $\gl_4$, we can use these formulae to evaluate 
the coefficients $B_k$ and $C_k$.

Evaluating (\ref{newT1}) at $v_k+2\eta$ and substituting the result into $B_k$,  we obtain
\bea
\la{Bev}
B_k=A_k+\frac{1}{\zt_1}\frac{Q_1(v_k+2\hb) }{Q_1(v_k+\hb)}\frac{\DP_1(v_k+\hb)}{\DP_1(v_k+2\hb)} \mathcal{A}(v_k+\hb)\, ,
\eea
where $\mathcal{A}(u)$ is defined in (\ref{Anew}). 

Similarly, evaluating (\ref{T2_res}) at $v_k+\eta$ and substituting the result into $C_k$, we find
\bea
\la{Cev}
C_k=
    \frac{\bar{\chi}_2(\zt)}{\zt_1^2} +\frac{1}{\zt_1}\frac{Q_1(v_k+2\hb)}{Q_1(v_k+\hb)}\frac{\DP_1(v_k+\hb)}{\DP_1(v_k+2\hb)}\mathcal{A}(v_k+\hb)
    -\frac{1}{\zt_1}\sum_{l=1}^{\mb_1}\frac{\hb \mathcal{C}_l \mathcal{A}(v_l)}{v_k-v_l+\hb}\, .
\eea
This derivation of the coefficients $B_k$ and $C_k$ avoids the use of resummation and straightforwardly delivers the coefficients that allow for a well-defined homogeneous limit.
Substituting (\ref{Dev}), (\ref{Bev}) and (\ref{Cev}) into  (\ref{ABCD}), $A_k$ and the terms proportional to $ \mathcal{A}(v_k+\hb)$ cancel and we obtain
\bea
\la{gl4cons}
 \frac{\bar{\chi}_2(\zt)}{\zt_1^2} + \frac{\chi_4(\zt)}{\zt_1^4} \frac{Q_1(v_k-\eta)}{Q_1(v_k+\eta)}\frac{\DP_1(v_k+\hb)}{\DP_1(v_k)} -\frac{1}{\zt_1}\sum_{l=1}^{\mb_1}\frac{\hb \mathcal{C}_l \mathcal{A}(v_l)}{v_k-v_l+\hb}=0\, .
\eea
Using the definition (\ref{Anew}) of $\mathcal{A}(u)$, this equation can be written as 
\bea
\la{newgl4}
&&\mathcal{R}_4:=\frac{\chi_4(\zt)}{\delta_1^4} \frac{Q_1(v_k-\eta)}{Q_1(v_k+\eta)}\frac{\DP_1(v_k+\hb)}{\DP_1(v_k)}\\
\nonumber
&&\hspace{1cm}+
 \frac{\bar{\chi}_2(\zt)}{\zt_1^2} -\frac{\bar{\chi}_1(\zt)}{\zt_1}\sum_{n=1}^{\mb_1}\frac{\hb \mathcal{C}_n}{v_k-v_n+\hb} +
\sum_{n_1=1}^{\mb_1}\sum_{n_2=1}^{\mb_1}\frac{\hb^2\mathcal{C}_{n_1}\mathcal{C}_{n_2}}{(v_k-v_{n_1}+\hb)(v_{n_1}-v_{n_2}+\hb)}=0\, ,
\eea
which provides a closed system of equations for determination of the principal Bethe roots $v_k$ of the $\gl_4$ spin chain in the vector representation.

For later comparison, it is convenient  to rewrite the $\gl_3$ equation (\ref{alm_B_2}) in the same notation:
\bea
\la{newgl3}
\mathcal{R}_3:=\frac{\chi_3(\delta)}{\zt_1^3}\frac{Q_1(v_k-\eta)}{Q_1(v_k+\eta)}\frac{\DP_1(v_k+\hb)}{\DP_1(v_k)}
+\frac{\bar{\chi}_1}{\zt_1}-\sum_{n=1}^{\mb_1}\frac{\eta \mathbcal{C}_n}{v_k-v_n+\hb}=0\, .
\eea
In both (\ref{newgl4}) and (\ref{newgl3}), the coefficients $\mathcal{C}_k$ are defined by (\ref{Anew}).
The characters appearing in (\ref{newgl4}) are those of $\gl_4$, whereas in (\ref{newgl3}) they are of $\gl_3$.
\medskip

\paragraph{\textsf{Embedding  property}.} \hspace{-0.2cm}We now show that every solution of the $\gl_3$ equations (\ref{newgl3}) provides a solution of the $\gl_4$ equations (\ref{newgl4}). 
Let $\{v_k\}$ satisfy $\mathcal{R}_3=0$. 
From (\ref{newgl3}) we get 
\bea
\la{newglN}
\sum_{n=1}^{\mb_1}\frac{\eta \mathbcal{C}_n}{v_k-v_n+\hb}=\frac{\zt_2\zt_3}{\zt_1^2}\gg_k
+\frac{\zt_2+\zt_3}{\delta_1}\, .
\eea
Here  $\gg_k=f(v_k)$ with $f(v)$ defined in (\ref{fv}).
Substitute (\ref{newglN}) this into the double sum of (\ref{newgl4}), we obatin
\bea
\nonumber
&&\sum_{n_1=1}^{\mb_1}\sum_{n_2=1}^{\mb_1}\frac{\hb^2\mathcal{C}_{n_1}\mathcal{C}_{n_2}}{(v_k-v_{n_1}+\hb)(v_{n_1}-v_{n_2}+\hb)}
=\sum_{n_1=1}^{\mb_1}\frac{\hb \mathcal{C}_{n_1}}{v_k-v_{n_1}+\hb}
\bigg(\frac{\zt_2\zt_3}{\zt_1^2}\gg_{n_1}
+\frac{\zt_2\zt_3}{\delta_1}\bigg)\\
\la{DS_ev}
&&\hspace{2cm}=\frac{\zt_2\zt_3}{\zt_1^2}\sum_{n=1}^{\mb_1}\frac{\hb\mathcal{C}_{n}\gg_{n}}{v_k-v_{n}+\hb}
+\frac{\zt_2+\zt_3}{\delta_1}\sum_{n=1}^{\mb_1}\frac{\eta \mathbcal{C}_n}{v_k-v_n+\hb}\, .
\eea
Using 
\bea
\la{Cngn}
\mathcal{C}_n=-\frac{1}{\gg_n}\prod_{l\neq n}^{\mb_1}\frac{v_n-v_l-\hb}{v_n-v_l}\, ,
\eea
together with the identity (\ref{RSident}),  equation (\ref{DS_ev}) reduces to
\bea
\la{DS_ex1}
\sum_{n_1=1}^{\mb_1}\sum_{n_2=1}^{\mb_1}\frac{\hb^2\mathcal{C}_{n_1}\mathcal{C}_{n_2}}{(v_k-v_{n_1}+\hb)(v_{n_1}-v_{n_2}+\hb)}=
-\frac{\zt_2\zt_3}{\zt_1^2}+\frac{\zt_2+\zt_3}{\delta_1}\sum_{n=1}^{\mb_1}\frac{\eta \mathbcal{C}_n}{v_k-v_n+\hb}\, .
\eea
Substituting (\ref{DS_ex1}) into (\ref{newgl4}) and using
\bea
\la{char4}
\bar{\chi}_1=\zt_2+\zt_3+\zt_4\, , ~~~~
\bar{\chi}_2=\zt_2\zt_3+\zt_4(\zt_2+\zt_3)\, , ~~~~
\chi_4=\zt_1\zt_2\zt_3\zt_4\, ,
\eea
we obtain
\bea
\mathcal{R}_4=\frac{\zt_4}{\zt_1}\bigg[ \frac{\zt_2\zt_3}{\zt_1^2}{\gg}_k +\frac{\zt_2+\zt_3}{\zt_1}-\sum_{n=1}^{\mb_1}\frac{\eta \mathbcal{C}_n}{v_k-v_n+\hb}\bigg]
=\frac{\zt_4}{\zt_1}\mathcal{R}_3=0\, .
\eea
Therefore every solution of the $\gl_3$ equations automatically satisfies the $\gl_4$ equations.
\medskip

The reduction from $\gl_4\to \gl_2$ can be obtained in a similar way. Substituting (\ref{Cngn}) into (\ref{newgl4}), we find
\bea
\la{newgl_mod}
&&\hspace{-0.3cm}\frac{\chi_4(\zt)}{\delta_1^4}\gg_k +
 \frac{\bar{\chi}_2(\zt)}{\zt_1^2} +\frac{\bar{\chi}_1(\zt)}{\zt_1}\sum_{n\neq l}^{\mb_1}\frac{1}{\gg_n}\frac{\hb }{v_k-v_n+\hb}\prod_{l=1}^{\mb_1}\frac{v_n-v_l-\hb}{v_n-v_l}
 \\
 \nonumber
&&\hspace{-0.3cm}+
\sum_{n_1=1}^{\mb_1}\sum_{n_2=1}^{\mb_1}\frac{1}{\gg_{n_1}\gg_{n_2}}
\frac{\hb^2}{(v_k-v_{n_1}+\hb)(v_{n_1}-v_{n_2}+\hb)}\!\prod_{l_1\neq n_1}^{\mb_1}\!\!\frac{v_{n_1}-v_{l_1}-\hb}{v_{n_1}-v_{l_1}}
\!\!\prod_{l_2\neq n_2}^{\mb_1}\!\!\frac{v_{n_2}-v_{l_2}-\hb}{v_{n_2}-v_{l_2}}=0\, .
\eea
Using the identity 
\bea
\hspace{-1cm}\sum_{n_1=1}^{\mb_1}\sum_{n_2=1}^{\mb_1}
\frac{\hb^2}{(v_k-v_{n_1}+\hb)(v_{n_1}-v_{n_2}+\hb)}\prod_{l_1\neq n_1}^{\mb_1}\frac{v_{n_1}-v_{l_1}-\hb}{v_{n_1}-v_{l_1}}
\prod_{l_2\neq n_2}^{\mb_1}\frac{v_{n_2}-v_{l_2}-\hb}{v_{n_2}-v_{l_2}}=1\, ,
\eea
equation (\ref{newgl_mod}) admits a constant solution $\gg_k=\gg$ for all $k=1,\ldots, k$. In this case  it reduces to 
\bea
\la{newgl_mod_const}
&&\frac{\chi_4(\zt)}{\zt_1^4}\gg^3+
 \frac{\bar{\chi}_2(\zt)}{\zt_1^2}\gg^2 +\frac{\bar{\chi}_1(\zt)}{\zt_1}\gg+1=0\, .
\eea
The roots of this cubic equation are 
\bea
\gg=-\frac{\zt_1}{\zt_2}\, , ~~~\gg=-\frac{\zt_1}{\zt_3}\, , ~~~\gg=-\frac{\zt_1}{\zt_4}.
\eea
These solutions correspond to the reduction of the  $\gl_4$ equations  to the $\gl_2$ Bethe equations, $\gg=f(v_k)$. 

\medskip 

\paragraph{\textsf{$\rm{U}(1)$ charges}.}\hspace{-0.2cm}Comparing the expression (\ref{BA_inspired}) for $\mT_1(u)$ 
in terms of Baxter's $Q$-polynomials with formula (\ref{newT1}), we find that 
\bea
\la{Agl4}
\mathcal{A}(u)=\zt_2\frac{Q_1(u)}{Q_1(u+\hb)}\frac{Q_2(u+2\hb)}{Q_2(u+\hb)}+\zt_3 \frac{Q_2(u)}{Q_2(u+\hb)}\frac{Q_3(u+2\hb)}{Q_3(u+\hb)}
+\zt_4\frac{Q_3(u)}{Q_3(u+\hb)}\, .
\eea
Similarly, comparing the Baxter-polynomial expression (\ref{BA_inspired})  for $\mT_2(u)$ with (\ref{T2_res}), while using  (\ref{Agl4}) for $\mathcal{A}(u)$, we find
\bea
\la{AAgl4}
\begin{aligned}
\bar{\chi}_2(\zt)-\zt_1\sum_{k=1}^{\mb_1}\frac{\hb  \mathcal{C}_k\mathcal{A}(v_k)}{u-v_k}&=
\zt_2\zt_3 \frac{Q_1(u-\hb)}{Q_1(u)} \frac{Q_3(u+2\hb)}{Q_3(u+\hb)}\\
&+\zt_2\zt_4  \frac{Q_1(u-\hb)}{Q_1(u)} \frac{Q_2(u+\hb)}{Q_2(u)} \frac{Q_3(u)}{Q_3(u+\hb)}
+\zt_3\zt_4\frac{Q_2(u-\hb)}{Q_2(u)}\, .
\end{aligned}
\eea
Expanding (\ref{Agl4}) and (\ref{AAgl4}) in the limit $u\to \infty$, the leading terms cancel, while the subleading ones yield 
\bea
\la{m2m3}
\begin{aligned}
&(\zt_2-\zt_3)\mb_2+(\zt_3-\zt_4)\mb_3=\zt_2 \mb_1 -\zt_1\sum_{k=1}^{\mb_1}\mathcal{C}_k\, ,\\
&\zt_4(\zt_2-\zt_3)\mb_2+\zt_2(\zt_3-\zt_4)\mb_3=\zt_2(\zt_3+\zt_4)\mb_1-\zt_1\sum_{k=1}^{\mb_1}\mathcal{C}_k\mathcal{A}(v_k)\, .
\end{aligned}
\eea
Evaluating $\mathcal{A}(v_k)$ using (\ref{Anew}), the system (\ref{m2m3}) can be solved for $m_2$ and $m_3$, yielding
\bea
\la{solm2m3}
\begin{aligned}
\mb_2=\frac{\zt_2(\zt_2-\zt_3-\zt_4)}
{(\zt_2-\zt_3)(\zt_2-\zt_4)}\mb_1 
&+
\frac{\zt_1(\zt_3+\zt_4)}{(\zt_2-\zt_3)(\zt_2-\zt_4)}\sum\limits_{k=1}^{\mb_1}\mathcal{C}_k \\
&-\frac{\zt_1^2   }
{(\zt_2-\zt_3)(\zt_2-\zt_4)}\sum\limits_{k,l=1}^{\mb_1}\frac{\hb\mathcal{C}_k\mathcal{C}_l}{v_k-v_l+\hb}\, , \\
\mb_3=\frac{\zt_2\zt_3}{(\zt_3-\zt_4)(\zt_2-\zt_4)}\mb_1
&-\frac{\zt_1(\zt_2+\zt_3)}{(\zt_3-\zt_4)(\zt_2-\zt_4)}\sum\limits_{k=1}^{\mb_1}\mathcal{C}_k\\
&+\frac{\zt_1^2}{(\zt_3-\zt_4)(\zt_2-\zt_4)}\sum\limits_{k,l=1}^{\mb_1}\frac{\hb\mathcal{C}_k\mathcal{C}_l}{v_k-v_l+\hb} \, .
\end{aligned}
\eea
Numerical investigations indicate that, for generic twists, the expressions  (\ref{solm2m3}) always yield integer values. 
They can be identified with the numbers $\mb_2$ and $\mb_3$ of 
auxiliary roots in the conventional nested Bethe ansatz corresponding to the solution $\{v_k\}$ of (\ref{newgl4}).
\medskip

\paragraph{\textsf{Eigenvalue of $\mT_3(u)$.}}\hspace{-0.2cm}To complete the analysis of the $\gl_4$ case, we derive the transfer-matrix eigenvalue $\mT_3(u)$ using the same 
interpolation and resummation procedure as 
for $\mT_1(u)$ and $\mT_2(u)$
\begin{align}
\notag
\mT_3(u)
&=
\DP_1(u+2\eta)\DP_1(u+3\eta)\\
&\times 
\left[
\zt_1\DP_1(u)\,
\frac{Q_1(u+\eta)}{Q_1(u)}
\,\mathcal{B}(u+\eta)
+
\bar{\chi}_3\,\DP_1(u+\eta)\,
\frac{Q_1(u-\eta)}{Q_1(u)}
\right]\, ,
\end{align}
where 
\begin{equation}
\la{Bu}
\mathcal{B}(u)
=
\bar{\chi}_2
-
\delta_1
\sum_{k=1}^{\mb_1}
\frac{\hb \mathcal{C}_k\,\mathcal{A}(v_k)}
     {u-v_k},
\end{equation}
while $\bar{\chi}_2=\zt_2\zt_3+\zt_2\zt_4+\zt_3\zt_4$ and $\bar{\chi}_3=\zt_2\zt_3\zt_4$. 
Requiring the cancellation of the apparent poles at $u=v_k$ reproduces the Bethe equations (\ref{gl4cons}). We come back to this important point in the next section.
The poles at $u=v_k-\hb$ originating from $\mathcal{B}(u+\hb)$ are removable, since they are multiplied by $Q_1(u+\hb)$, which vanishes at the same points. 
Consequently, $\mT_3(u)$ is a polynomial on the solution space of $\mathcal{R}_4=0$. Furthermore, $\mT_3(u)\sim
\chi_3^{(4)}(\zt)u^{3N}$, $u\to \infty$, in agreement with the expected asymptotic behaviour.

\section{Generalization to Arbitrary Rank}
\paragraph{\textsf{Equations for arbitrary rank.}}\hspace{-0.2cm}To generalize the equations we obtained for the $\gl_3$ and $\gl_4$ cases, we introduce the following notation for the reduced twist
\bea
\la{RT}
\Delta^{(\ell)}_{\xx}=\sum_{\a=0}^\xx (-1)^{\xx-\a}\frac{\chi^{(\ell)}_{\raisebox{-0.4ex}{$\scriptstyle \a$}}(\zt)}{\zt_1^\a}\, .
\eea
Then the natural generalization of the above cases to the $\gl_{\ell}$ is given by 
\begin{align}
\notag
\mathcal{R}_{\ell}:&= \Delta^{(\ell)}_{\ell-2}+\Delta_{\ell-1}^{(\ell)}\frac{Q_1(v_k-\eta)}{Q_1(v_k+\eta)}\frac{\DP_1(v_k+\hb)}{\DP_1(v_k)}\\
    &+\sum_{\xx=1}^{\ell-2}(-1)^\xx\Delta^{(\ell)}_{\ell-2-\xx}
   \sum_{n_1,\ldots, n_\xx =1}^{\mb_1}\frac{\hb \mathcal{C}_{n_1}}{v_k-v_{n_1}+\hb}
    \prod_{j=2}^\xx \frac{\hb \mathcal{C}_{n_j} }{v_{n_{j-1}}-v_{n_{j}}+\hb}=0\, ,\label{conjecture}
\end{align}
where the coefficients $\mathcal{C}_k$ are defined in (\ref{Anew}). 
Formally, the term $\Delta^{(\ell)}_{\ell-2}$ 
can be incorporated into the summation by extending it to $\xx=0$.

\medskip
Clearly, for $\ell=3$ and $\ell=4$ equations (\ref{conjecture}) reproduce (\ref{newgl3}) and (\ref{newgl4}), respectively.
For $\ell=2$ the entire summation over $r$ is absent. Using the explicit values of $\Delta_0^{(2)}$ and $\Delta_1^{(2)}$,
we obtain 
\bea
\la{sl2rec}
\frac{Q_1(v_k-\hb)}{Q_1(v_k+\hb)}=-\frac{\zt_1}{\zt_2}\frac{\DP_1(v_k)}{\DP_1(v_k+\hb)}\, .
\eea
After the shift $v_k\to v_k+\hb/2$, equation (\ref{sl2rec}) coincides with the standard twisted $\gl_2$ Bethe equations written in Baxter-polynomial form.
Thus, the hierarchy naturally originates from the $\gl_2$ case.

Using the inverse of the Cauchy kernel 
\bea
\la{Cauch_K}
K_{kn}=\frac{\hb \mathcal{C}_n}{v_k-v_n+\hb}\, ,
\eea 
equation (\ref{conjecture}) can be rewritten in the equivalent dual form 
\begin{align}
\notag
\widetilde{\mathcal R}_{\ell}:&=\frac{Q_1(v_k+\eta)}{Q_1(v_k-\eta)}\frac{\DP_1(v_k)}{\DP_1(v_k+\hb)}\\
&+
\sum_{\xx=0}^{\ell-2}(-1)^\xx \Delta_{\xx+1}^{(\ell)} \sum_{n_1,\ldots, n_\xx =1}^{\mb_1}\frac{\hb \widetilde{\mathcal C}_{n_1}}{v_k-v_{n_1}-\hb}
    \prod_{j=2}^{\xx} \frac{\hb \widetilde{\mathcal C}_{n_j} }{v_{n_{j-1}}-v_{n_{j}}-\hb}=0\, ,
    \la{dual}
\end{align}
where the coefficients $\widetilde{\mathcal C}_{n}$ are given by
\bea
\widetilde{\mathcal C}_{n}=-\frac{\DP_1(v_n+\hb)}{\DP_1(v_n)}\prod_{s\neq n}^{\mb_1}\frac{v_n-v_s-\hb}{v_n-v_s}\, .
\eea

\medskip
\vspace{-0.3cm}
\paragraph{\textsf{Embedding property}.}\hspace{-0.2cm}A useful consistency check of the conjectured equations (\ref{conjecture}) is the embedding property with respect to the rank.
More precisely, every solution of 
 $\mathcal{R}_{\ell-1}$ also solves $\mathcal{R}_{\ell}$.  
 Let $\{v_n\}_{n=1}^{\mb_1}$ be a solution of the rank-$(\ell-1)$ equations, i.e.,
 \bea
 \la{conj_l}
 \mathcal{R}_{\ell-1}(v_n)=0 , ~~~~n=1,\ldots, \mb_1\, ,
 \eea 
where 
\begin{align}
\notag
\mathcal{R}_{\ell-1}(v_n)&= \Delta^{(\ell-1)}_{\ell-3}+\Delta_{\ell-2}^{(\ell-1)}\frac{Q_1(v_n-\eta)}{Q_1(v_n+\eta)}\frac{\DP_1(v_n+\hb)}{\DP_1(v_n)}\\
    &+\sum_{\xx =1}^{\ell-3}(-1)^\xx \Delta^{(\ell-1)}_{\ell-3-\xx}
   \sum_{n_1,\ldots, n_\xx=1}^{\mb_1} \frac{\hb \mathcal{C}_{n_1}}{v_n-v_{n_1}+\hb}
    \prod_{j=2}^\xx \frac{\hb \mathcal{C}_{n_j} }{v_{n_{j-1}}-v_{n_{j}}+\hb}\, .\label{conjecture_1}
\end{align}
Multiply (\ref{conj_l})  by 
\bea
\frac{\hb\mathcal{C}_n}{v_k-v_n+\hb}
\eea
and summing over $n$, we make use of the identity
\bea
\sum_{n=1}^{\mb_1}\frac{\hb\mathcal{C}_n}{v_k-v_n+\hb}\frac{Q_1(v_n-\eta)}{Q_1(v_n+\eta)}\frac{\DP_1(v_n+\hb)}{\DP_1(v_n)}=-1\, ,
\eea
to obtain
\begin{align}
\notag
-\Delta_{\ell-2}^{(\ell-1)}&+ \Delta^{(\ell-1)}_{\ell-3}\sum_{n=1}^{\mb_1}\frac{\hb\mathcal{C}_n}{v_k-v_n+\hb}\\
&+\sum_{\xx=1}^{\ell-3}(-1)^\xx\Delta^{(\ell-1)}_{\ell-3-\xx}
   \sum_{n_1,\ldots, n_{\xx +1}=1}^{\mb_1}\frac{\hb \mathcal{C}_{n_1}}{v_k-v_{n_1}+\hb}
    \prod_{j=2}^{\xx+1} \frac{\hb \mathcal{C}_{n_j} }{v_{n_{j-1}}-v_{n_{j}}+\hb}=0\, .\label{conjecture_2}
\end{align}
After shifting  the summation index $\xx\to \xx-1$, equation (\ref{conjecture_2}) becomes
\bea\la{imp_iden}
\Delta_{\ell-2}^{(\ell-1)}+\sum_{\xx=1}^{\ell-2}(-1)^\xx\Delta^{(\ell-1)}_{\ell-2-\xx}S_\xx(v_k)=0\, ,
\label{conjecture_3}
\eea
where 
\bea
\la{S_r}
S_\xx(v_k)=   \sum_{n_1,\ldots, n_\xx=1}^{\mb_1}\frac{\hb \mathcal{C}_{n_1}}{v_n-v_{n_1}+\hb}
    \prod_{j=2}^{\xx} \frac{\hb \mathcal{C}_{n_j} }{v_{n_{j-1}}-v_{n_{j}}+\hb}\, .
\eea
For future convenience we also introduce 
\bea
\la{S_rt}
\widetilde{S}_\xx(v_k)=   \sum_{n_1,\ldots, n_\xx=1}^{\mb_1}\frac{\hb \widetilde{\mathcal C}_{n_1}}{v_n-v_{n_1}-\hb}
    \prod_{j=2}^{\xx} \frac{\hb \widetilde{\mathcal C}_{n_j} }{v_{n_{j-1}}-v_{n_{j}}-\hb}\, .
\eea
We further note that the reduced twists satisfy the following recurrence relations\footnote{The relation $\Delta_\xx^{(\xx)}=0$ follows from the characteristic polynomial identity\\
$~$\hspace{0.56cm}$\sum\limits_{\a=0}^\xx(-1)^{\xx-\a}\frac{\chi_\a^{(\xx)}}{\zt_1^\a}=\frac{1}{\zt_1^\xx}\prod\limits_{\a=1}^\xx (\zt_\a-\zt_1)=0$.}
\bea
\Delta_\xx^{(\ell)}=\Delta_\xx^{(\ell-1)}+\frac{\zt_{\ell}}{\zt_1}\Delta_{\xx-1}^{(\ell-1)}\, ,~~~~~\Delta_\xx^{(\xx)}=0\, .
\eea 
In particular, $\Delta_{\ell-1}^{(\ell)}=\frac{\zt_{\ell}}{\zt_1}\Delta_{\ell-2}^{(\ell-1)}$.
Substituting these relations into (\ref{conjecture}), we obtain
\begin{align}
\notag
\mathcal{R}_{\ell}= \Delta_{\ell-2}^{(\ell-1)}&+\frac{\zt_{\ell}}{\zt_1}\Delta_{\ell-3}^{(\ell-1)}+\frac{\zt_{\ell}}{\zt_1}\Delta_{\ell-2}^{(\ell-1)}\frac{Q_1(v_k-\eta)}{Q_1(v_k+\eta)}\frac{\DP_1(v_k+\hb)}{\DP_1(v_k)}\\
    &+\sum_{\xx=1}^{\ell-2}(-1)^\xx\Delta^{(\ell-1)}_{\ell-2-\xx} S_\xx(v_k)+\frac{\zt_{\ell}}{\zt_1}\sum_{\xx=1}^{\ell-2}(-1)^\xx\Delta^{(\ell-1)}_{\ell-3-\xx}S_\xx(v_k)\, .\label{conjecture_x}
\end{align}
Rearranging the terms yields
\begin{align}
\label{conjecture_xx}
\mathcal{R}_{\ell}= \bigg[\Delta_{\ell-2}^{(\ell-1)}&+\sum_{\xx=1}^{\ell-2}(-1)^\xx\Delta^{(\ell-1)}_{\ell-2-\xx} S_\xx(v_k)\bigg]\\
&+\frac{\zt_{\ell}}{\zt_1}\bigg[\Delta_{\ell-3}^{(\ell-1)}+\Delta_{\ell-2}^{(\ell-1)}\frac{Q_1(v_k-\eta)}{Q_1(v_k+\eta)}\frac{\DP_1(v_k+\hb)}{\DP_1(v_k)}
    +\sum_{\xx=1}^{\ell-2}(-1)^\xx\Delta^{(\ell-1)}_{\ell-3-\xx}S_\xx(v_k)\bigg]\, .
\notag
\end{align}
The expression in the second bracket coincides with $\mathcal{R}_{\ell-1}$, while the first bracket vanishes by virtue of  (\ref{imp_iden}). 
Therefore 
\bea
\mathcal{R}_{\ell}=\frac{\zt_{\ell}}{\zt_1}\mathcal{R}_{\ell-1}=0\, .
\eea
Thus every solution of $\mathcal{R}_{\ell-1}$ is also a solution of $\mathcal{R}_{\ell}=0$. This confirms the hierarchical embedding
\bea
\gl_2\subset \gl_3\subset \ldots \subset \gl_{\ell}\, .
\eea

We have also compared numerical solutions of (\ref{conjecture}) for the $\gl_5$ spin chain
with those obtained from the conventional nested Bethe ansatz. 
Complete agreement was found in all cases considered.

\paragraph{\textsf{Untwisted and homogeneous limits}.}\hspace{-0.2cm}In the untwisted limit $\zt_1=\ldots=\zt_{\ell}=1$, the reduced twist coefficients become
\bea
\la{Duntw}
\Delta_\xx^{(\ell)}=\binom{\ell-1}{\xx}\, .
\eea
These coefficients are precisely the characters of the fundamental representation of the reduced algebra $\gl_{\ell-1}$. Substituting (\ref{Duntw}) into (\ref{conjecture})
and shifting all Bethe roots $v_k\to v_k+\hb/2$, we obtain
\bea
\la{untwistR}
\mathcal{R}^{\rm untw}_{\ell}=
\frac{Q_1(v_k-\eta)}{Q_1(v_k+\eta)}\frac{\DP(v_k+\hb/2)}{\DP(v_k-\hb/2)}+\sum_{\xx=0}^{\ell-2}(-1)^\xx \binom{\ell-1}{\xx+1}S_\xx(v_k)=0\, .
\eea
Here the summation extends to $\xx=0$ with the convention $S_0(v_k)=1$ and 
the coefficients $\mathcal{C}_n$ entering $S_\xx(v_k)$ for $r\geq 1$ become
\bea
\la{untwistCn}
\mathcal{C}_n=\frac{\DP(v_n-\hb/2)}{\DP(v_n+\hb/2)}\prod_{s\neq n}^{\mb_1}\frac{v_n-v_s+\hb}{v_n-v_s}\, ,
\eea
Here we introduced the polynomial 
\bea
\la{weta}
\DP(u)=\prod_{j=1}^N (u-u_j)\, .
\eea
Applying the same shift to the dual representation  (\ref{dual}), we obtain for the untwisted case  
\bea
\la{untwistRdual}
\widetilde{\mathcal R}^{\rm untw}_{\ell}=\frac{Q_1(v_k+\eta)}{Q_1(v_k-\eta)}\frac{\DP(v_k-\hb/2)}{\DP(v_k+\hb/2)}+\sum_{\xx=0}^{\ell-2}(-1)^\xx \binom{\ell-1}{\xx+1}\widetilde{S}_\xx(v_k)=0\, ,
\eea
where the coefficients $\tilde{\mathcal C}_n$ in $\widetilde{S}_\xx(v_k)$ become 
\bea
\la{untwistCndual}
\widetilde{\mathcal C}_{n}=-\frac{\DP(v_n+\hb/2)}{\DP(v_n-\eta/2)}\prod_{s\neq n}^{\mb_1}\frac{v_n-v_s-\hb}{v_n-v_s}\, .
\eea
Both (\ref{untwistR}) and (\ref{untwistRdual}) admit a well-defined homogeneous limit $u_j\to 0$.

\paragraph{\textsf{Conjugation}.}\hspace{-0.2cm}We now establish the conjugation property of the untwisted equations.
Throughout this discussion we assume
\bea
u_j\in\mathbb R,\qquad
\eta^*=-\eta,
\eea
while all Bethe roots are understood to be the shifted variables introduced
above.

Under these assumptions the driving terms satisfy
\bea
\left(
\frac{Q_1(v_k-\eta)}{Q_1(v_k+\eta)}
\frac{\DP(v_k+\eta/2)}{\DP(v_k-\eta/2)}
\right)^*
=
\frac{Q_1(v_k^*+\eta)}{Q_1(v_k^*-\eta)}
\frac{\DP(v_k^*-\eta/2)}{\DP(v_k^*+\eta/2)},
\eea
which is precisely the driving term of the dual equation
(\ref{untwistRdual}).
Furthermore, using (\ref{untwistCn}) and (\ref{untwistCndual}), one finds
\bea
{\mathcal C}_n(v)^*
=
-\widetilde{\mathcal C}_n(v^*),
\eea
where the product is evaluated on the complex-conjugated Bethe roots. Since $\hb^*=-\hb$, the combinations entering the sums obey $\big(\hb{\mathcal C}_n(v)\big)^*
=
\hb\widetilde{\mathcal C}_n(v^*),$
Consequently,
\bea
S_\xx(v_k)^*
=
\widetilde S_\xx(v_k^*).
\eea
Taking the complex conjugate of (\ref{untwistR}) therefore yields
\bea
\frac{Q_1(v_k^*+\eta)}{Q_1(v_k^*-\eta)}
\frac{\DP(v_k^*-\eta/2)}{\DP(v_k^*+\eta/2)}
+
\sum_{\xx=0}^{\ell-2}
(-1)^\xx
\binom{\ell-1}{\xx+1}
\widetilde S_\xx(v_k^*)
=0,
\eea
which is exactly the dual equation (\ref{untwistRdual}) evaluated on the
complex-conjugated roots.

Applying the Cauchy inversion procedure to the dual equation (\ref{untwistRdual}) reproduces the original
equations (\ref{untwistR}). Therefore, complex conjugation interchanges the original and  and dual systems. Consequently, if
\(
\{v_1,\ldots,v_{\mb_1}\}
\)
solves the untwisted Bethe equations, so does
\(
\{v_1^*,\ldots,v_{\mb_1}^*\}.
\)
Hence, for real inhomogeneities and purely imaginary $\hb$, the untwisted Bethe equations are invariant under complex conjugation, 
\bea
\{v_1,\ldots,v_{\mb_1}\}\ \longmapsto\
\{v_1^*,\ldots,v_{\mb_1}^*\}.
\eea
We have shown that complex conjugation maps every solution of the untwisted equations to another solution. This establishes the equivariance of the solution set under complex conjugation. For hermitian spin chains with real inhomogeneities, one expects each physical solution to be self-conjugate, in agreement with the conventional string hypothesis in the thermodynamic limit. Establishing this stronger property requires additional spectral input beyond the Bethe equations themselves.

Having established the untwisted and homogeneous limits of the conjectured equations (\ref{conjecture}), together with their conjugation properties in the untwisted case, 
we now return to the transfer-matrix description 
and discuss a possible organizing principle underlying the corresponding eigenvalues. 
\medskip

\paragraph{\textsf{Transfer matrices and regularity conditions.}}\hspace{-0.2cm}The explicit expressions  for $\mT_1(u)$, $\mT_2(u)$ and $\mT_3(u)$ obtained in the previous sections 
suggest a recursive structure that may persist for arbitrary rank. Introducing a sequence of functions $\Phi_\xx(u)$ through
\bea
\la{recurs}
\begin{aligned}
\Phi_0(u)&=1, \\
\Phi_\xx(u)&=
\zt_1^\xx \Delta_\xx^{(\ell)}
-\zt_1
\sum_{n=1}^{\mb_1}
\frac{\eta \mathcal{C}_n\,\Phi_{\xx-1}(v_n+\hb)}
     {u-v_n},
\qquad
\xx\geq 1\, ,
\end{aligned}
\eea
we are led to conjecture the representation
\begin{align}
\mT_\xx(u)=
g_\xx(u)
\left[
\zt_1 \DP_1(u)\frac{Q_1(u+\eta)}{Q_1(u)}
\,\Phi_{\xx-1}(u+\eta)
+
\DP_1(u+\eta)\Phi_\xx(u)
\right],
\label{Tk-conjecture}
\end{align}
valid for $\xx=1,\ldots,\ell-1$. Here $g_\xx(u)$ is defined in (\ref{def_g}).

In the vacuum sector, $\mb_1=0$, one has $Q_1(u)=1$ and $\Phi_0(u)=1$. The recursion (\ref{recurs}) therefore gives
\bea
\Phi_\xx(u)=\zt_1^\xx\Delta_\xx^{(\ell)},\qquad \xx\ge1,
\eea
and the conjectured representation (\ref{Tk-conjecture}) reduces to
\bea
\mT_\xx^{\rm vac}(u)
=
\zt_1^\xx g_\xx(u)
\Bigl[
\Delta_{\xx-1}^{(\ell)}\DP_1(u)
+
\Delta_\xx^{(\ell)}\DP_1(u+\eta)
\Bigr].
\eea
This agrees with the vacuum eigenvalues obtained from the tableau-sum formula (\ref{BA_inspired}).

For $\xx=1$ this formula reproduces
\begin{equation}
\la{Phi1}
\Phi_1(u)=\mathcal A(u-\hb)
=
\bar{\chi}_1
-\zt_1
\sum_{n=1}^{\mb_1}
\frac{\eta \mathcal{C}_n}{u-v_n+\eta},
\end{equation}
while for $\xx=2$ it yields
\begin{equation}
\la{Phi2}
\Phi_2(u)=\mathcal B(u)
=
\bar{\chi}_2
-\zt_1
\sum_{n=1}^{\mb_1}
\frac{\eta  \mathcal{C}_n\,\mathcal A(v_n)}
     {u-v_n}.
\end{equation}
Furthermore, in the $\gl_4$ case the relation
\begin{equation}
\la{Phi3}
\Phi_3(u)
=
\bar{\chi}_3
\frac{Q_1(u-\eta)}{Q_1(u)}
\end{equation}
holds on solutions of the Bethe equations, and (\ref{Tk-conjecture}) reproduces the explicit expression obtained for $\mT_3(u)$. 

The relation (\ref{Phi3}) differs qualitatively from the expressions (\ref{Phi1}) and (\ref{Phi2}). While $\Phi_1(u)$ and $\Phi_2(u)$ are obtained recursively as rational functions built from the coefficients $\mathcal{C}_n$, the highest function $\Phi_3(u)$ collapses on shell to a simple ratio of Baxter polynomials. This is not accidental. Indeed, for $\gl_4$ the function $\Phi_3$ corresponds to the last node of the hierarchy, and equation (\ref{Phi3}) is equivalent to the Bethe equations (\ref{gl4cons}). In this sense, $\Phi_3$ is not an independent object but is determined by the requirement that the transfer matrix $\mT_3(u)$ be polynomial.

The explicit $\gl_4$ result suggests that the same phenomenon persists for arbitrary rank. More precisely, one may expect that the functions
\bea
\Phi_0,\Phi_1,\ldots,\Phi_{\ell-2}
\eea
are generated recursively by (\ref{recurs}), whereas the final function is fixed by the Bethe equations and takes the universal form
\begin{equation}
\Phi_{\ell-1}(u)=
\zt_1^{\ell-1}\Delta_{\ell-1}^{(\ell)}
\frac{Q_1(u-\eta)}{Q_1(u)} .
\label{highest_Phi}
\end{equation}
For $\ell=4$, equation (\ref{highest_Phi}) reduces precisely to (\ref{Phi3}). Furthermore, using $\DP_1(u_j+\hb)=0$, the relation
$g_{\xx+1}(u_j)=\DP_1(u_j+2\hb)g_\xx(u_j+\hb)$, and the closure formula (\ref{highest_Phi}), one verifies that the conjectured representation
(\ref{Tk-conjecture})  is consistent with the fusion relations  (\ref{spec_sys_sep}), including the highest one,
$\mT_1(u_j)\mT_{\ell-1}(u_j)=\mT_{\ell}(u_j)$. 

Viewed from this perspective, the hierarchy
\bea
\Phi_0
\longrightarrow
\Phi_1
\longrightarrow
\cdots
\longrightarrow
\Phi_{\ell-2}
\longrightarrow
\delta_1^{\ell-1}\Delta_{\ell-1}^{(\ell)}
\frac{Q_1(u-\eta)}{Q_1(u)}
\eea
resembles a truncated $Q$-system.
The intermediate functions are generated recursively from the coefficients $\mathcal{C}_n$, while the last node is closed by the Bethe equations themselves. If this picture is correct, the cancellation of the apparent poles of the highest transfer matrix $\mT_{\ell-1}(u)$ should reproduce the rank-$\ell$ Bethe equations $\mathcal R_\ell=0$, providing a natural explanation for the structure of the conjectured hierarchy.  We now show that this is indeed the case.

The truncated $Q$-system picture also suggests a deeper interpretation of the hierarchy.
Rather than being introduced independently, the functions $\Phi_\xx$ appear to be generated by the requirement 
that the transfer matrices $\mT_\xx(u)$ remain polynomial. 
 Indeed, for $\xx=1,\ldots,\ell-2,$ all
apparent poles of $\mT_\xx(u)$ are located at the zeros $u=v_n$ of the Baxter
polynomial $Q_1(u)$. These poles originate both from the explicit factor $1/Q_1(u)$
and from the poles of $\Phi_\xx(u)$, which occur at the same locations.\footnote{
The poles of $\Phi_{\xx-1}(u+\eta)$ occur at $u=v_n-\eta$ and are
cancelled by the accompanying factor $Q_1(u+\eta)$. Therefore the only poles whose
cancellation produces independent constraints are those at $u=v_n$.
}
Omitting the common prefactor in  (\ref{Tk-conjecture}),  the residue of the first term at $u=v_n$ is 
\bea
\zt_1 \DP_1(v_n)
\frac{Q_1(v_n+\eta)}{Q_1'(v_n)}
\Phi_{\xx-1}\!\bigl(v_n+\eta\bigr).
\eea
The residue of the second term at the same point is
\bea
-\DP_1(v_n+\hb)\zt_1 \eta \mathcal{C}_n\,
\Phi_{\xx-1}\!\bigl(v_n+\eta\bigr)\, .
\eea
Using the definition of $\mathcal{C}_n$ (cf. (\ref{Anew})),
\bea
\eta \mathcal{C}_n
=
\frac{\DP_1(v_n)}{\DP_1(v_n+\eta)}
\frac{Q_1(v_n+\eta)}{Q_1'(v_n)}\, ,
\eea
one finds that two contributions cancel exactly. Equivalently, the cancellation of apparent poles at $u=v_n$
is precisely encoded in the recursive definition (\ref{recurs}). 
Thus the regularity conditions of
the ransfer matrices $\mT_\xx(u)$, $\xx=1,\ldots, \ell-2$, generate the hierarchy
of functions $\Phi_\xx$, while the regularity condition of 
$\mT_{\ell-1}(u)$ provides its closure. 

We now turn to the transfer matrix $\mT_{\ell-1}(u)$. Since $\mT_{\ell}(u)$ is an eigenvalue of the quantum determinant,
$\mT_{\ell-1}$ is the last nontrivial element of the hierarchy. We will show that its regularity condition
reproduces precisely the rank-$\ell$ Bethe equations $\mathcal{R}_{\ell}=0$. 
Indeed, the only nontrivial poles of $\mT_{\ell-1}(u)$ occur at the zeros $u=v_k$ of the Baxter polynomial $Q_1(u)$; the shifted poles are removable due to the accompanying zeroes 
of $Q_1(u+\hb)$. Requiring the cancellation of these apparent poles yields
\begin{equation}
\zt_1 \DP_1(v_k)Q_1(v_k+\eta)
\Phi_{\ell-2}(v_k+\eta)
+
\bar{\chi}_{\ell-1}
\DP_1(v_k+\eta)Q_1(v_k-\eta)
=0 .
\end{equation}
Using
\begin{equation}
\bar{\chi}_{\ell-1}=
\zt_1^{\ell-1}\Delta_{\ell-1}^{(\ell)},
\end{equation}
and dividing by
$\zt_1^{\ell-1}\DP_1(v_k)Q_1(v_k+\eta)$,
we obtain
\begin{equation}
F_{\ell-2}(v_k)
+
\Delta_{\ell-1}^{(\ell)}
\frac{Q_1(v_k-\eta)}{Q_1(v_k+\eta)}
\frac{\DP_1(v_k+\eta)}{\DP_1(v_k)}
=0 ,
\label{pole-cancellation}
\end{equation}
where we have introduced
\begin{equation}
F_\mu(v_k):=
\frac{1}{\zt_1^\mu}
\Phi_\mu(v_k+\eta).
\end{equation}
The recursion relation for $\Phi_\mu$ gives
\begin{equation}
\la{Fsv}
F_\mu(v_k)=
\Delta_\mu^{(\ell)}-
\sum_{n=1}^{\mb_1}
\frac{\eta \mathcal{C}_n}{v_k-v_n+\eta}
F_{\mu-1}(v_n).
\end{equation}
Iterating this relation, one finds
\begin{equation}
\la{Fhsol}
F_\mu(v_k)=
\Delta_\mu^{(\ell)}
+
\sum_{\xx=1}^{\mu}
(-1)^\xx
\Delta_{\mu -\xx}^{(\ell)}
S_\xx(v_k),
\end{equation}
where $S_\xx$ is defined in (\ref{S_r}).
Setting $\mu=\ell-2$ and substituting the resulting expression into \eqref{pole-cancellation}, we arrive at
\begin{equation}
\Delta_{\ell-2}^{(\ell)}
+
\Delta_{\ell-1}^{(\ell)}
\frac{Q_1(v_k-\eta)}{Q_1(v_k+\eta)}
\frac{\DP_1(v_k+\eta)}{\DP_1(v_k)}
+
\sum_{\xx=1}^{\ell-2}
(-1)^\xx
\Delta_{\ell-2-\xx}^{(\ell)}
S_\xx(v_k)
=0 .
\end{equation}
This equation coincides precisely with $\mathcal R_\ell(v_k)=0$. Therefore, the cancellation of the apparent poles of $\mT_{\ell-1}(u)$ reproduces the Bethe equations (\ref{conjecture}).


\section{Gaudin Limit and Relation to Scalar Opers}
It is natural to investigate the quasi-classical (Gaudin) limit of the non-nested Bethe equations and its relation to the oper description of the Gaudin model \cite{Frenkel2007,FrenkelBook,MTVOpers,MTV2009}.
For this purpose, we introduce the scaled twist parameters 
$
\zt_\a=\exp(\hb \upmu_{\a})$. Then, we have 
\bea
\Delta^{(\ell)}_{\xx}=\sum_{r=0}^\infty \frac{\hb^r}{r!}\sum_{1\leq \a_1<\ldots <\a_{\xx}\leq \ell-1} (\uptau_1+\ldots +\uptau_{\a_{\xx}})^r\, ,
\eea
where $\uptau_\a=\upmu_{\a+1}-\upmu_1$, $\a=1,\ldots, \ell-1$.  
The Gaudin model is obtained by taking the limit $\hb\to 0$
with $\upmu_\a$ kept fixed. 

For simplicity, we discuss explicitly only the $\gl_3$ case.
For $\ell=3$, the reduced twist coefficients expand as 
\bea
\begin{aligned}
\Delta^{(3)}_1
&=
2+\eta(\uptau_1+\uptau_2)+\frac{\hb^2}{2}(\uptau_1^2+\uptau_2^2)+O(\eta^2), \\
\Delta^{(3)}_2
&=
1+\eta(\uptau_1+\uptau_2)+\frac{\hb^2}{2}(\uptau_1+\uptau_2)^2+O(\eta^2),
\end{aligned}
\eea
where
$\uptau_1=\upmu_2-\upmu_1$,
and
$\uptau_2=\upmu_3-\upmu_1$.
Introduce the quantities 
\bea
\textsf{P}_k=\frac{\DP'(v_k)}{\DP(v_k)},\qquad
\textsf{V}_k=\sum_{n\neq k}^{\mb_1}\frac{1}{v_k-v_n},
\qquad
\textsf{W}_k=\sum_{n\neq k}^{\mb_1}\frac{1}{(v_k-v_n)^2}\, , 
\eea
where $\DP(u)$ is defined by (\ref{weta}). Then
\bea
\mathcal{C}_k
=
1+\eta(\textsf{V}_k-\textsf{P}_k)+\hb^2 \bigg[\frac{1}{2}(\textsf{V}_k-\textsf{P}_k)^2-\frac{1}{2}\textsf{W}_k\bigg]
+\mathcal{O}(\eta^3),
\eea
which in turn leads to 
\bea
S_1(v_k)
=
1-\eta \textsf{A}_k+\eta^2\textsf{B}_k+\mathcal{O}(\eta^3).
\eea
Here
\bea
\begin{aligned}
\textsf{A}_k
&=
\textsf{P}_k-2 \textsf{V}_k\, , \\
\textsf{B}_k&=
\frac12(\textsf{V}_k-\textsf{P}_k)^2-\frac32 \textsf{W}_k
+\sum_{n\neq k}^{\mb_1}\frac{\textsf{V}_n-\textsf{P}_n}{v_k-v_n}.
\end{aligned}
\eea
Substituting these expansions into the non-nested equation $\mathcal{R}_3(v_k)=0$, together with the expansion of the driving term 
\bea
\nonumber
&& \Delta^{(3)}_{1}+\Delta_{2}^{(3)}\frac{Q_1(v_k-\eta)}{Q_1(v_k+\eta)}\frac{\DP_1(v_k+\hb)}{\DP_1(v_k)}\\
&&\hspace{1cm} =1-\hb \textsf{A}_k-\hb^2
\bigg[\frac{1}{2}\textsf{A}_k^2+(\uptau_1+\uptau_2)\textsf{A}_k+\uptau_1\uptau_2
 \bigg]+\mathcal{O}(\hb^3)\, ,
\eea
 we find that the terms of order $\eta^0$ and $\eta^1$ cancel identically. The
leading non-trivial contribution is therefore of order $\eta^2$,
\bea
\mathcal{R}_3(v_k)
=
-\eta^2\mathcal{R}^{\rm G}_3(v_k)
+
\mathcal{O}(\hb^3),
\eea
where
\bea
\mathcal{R}^{\rm G}_3(v_k)
=
\textsf{B}_k
+
(\uptau_1+\uptau_2)\textsf{A}_k
+
\frac12\textsf{A}_k^2
+
\uptau_1\uptau_2\, .
\eea
Employing the identity 
\bea
\sum_{n\neq k}^{\mb_1}\frac{\textsf{V}_n}{v_k-v_n}=\frac{1}{2}\textsf{V}_k^2-\frac{3}{2}\textsf{W}_k\, ,
\eea
we rewrite the equation $\mathcal{R}^{\rm G}_3(v_k)=0$ in the form 
\bea
\hspace{-0.7cm}3\textsf{V}_k^2-3\textsf{W}_k-\big(3\textsf{P}_k+2(\uptau_1+\uptau_2)\big)\textsf{V}_k+\textsf{P}^2_k+(\uptau_1+\uptau_2)\textsf{P}_k+
\uptau_1\uptau_2
-\sum_{n\neq k}^{\mb_1}\frac{\textsf{P}_n}{v_k-v_n}=0\, .
\eea
The derivatives of the Baxter polynomial $Q_1(u)$ evaluated at a root $v_k$ satisfy
\bea
\frac{Q''_1(v_k)}{Q'_1(v_k)}
=2\textsf{V}_k\, , ~~~~
\frac{Q'''_1(v_k)}{Q'_1(v_k)}
=3\textsf{V}_k^2-3\textsf{W}_k\, .
\eea
We further use the identity
\bea
\sum_{n\neq k}^{\mb_1}\frac{\textsf{P}_n}{v_k-v_n}=\textsf{P}_k\textsf{V}_k+\textsf{P}'_k - \sum_{j=1}^N \frac{Q'_1(u_j)}{Q_1(u_j)}\frac{1}{v_k-u_j}\, ,
\eea
which follows from the residue theorem applied to an appropriate rational function.
Therefore, $\mathcal{R}^{\rm G}_3(v_k)=0$ becomes the Gaudin-type system
\bea
\la{no_pole}
\begin{aligned}
\frac{Q'''_1(v_k)}{Q'_1(v_k)}&-\big(2\textsf{P}(v_k)+\uptau_1+\uptau_2\big)\frac{Q''_1(v_k)}{Q'_1(v_k)}\\\
&+(\textsf{P}(v_k)+\uptau_1)(\textsf{P}(v_k)+\uptau_2)-\textsf{P}'(v_k)+
 \sum_{j=1}^N \frac{Q'_1(u_j)}{Q_1(u_j)}\frac{1}{v_k-u_j}=0\, ,
 \end{aligned}
\eea
where $\textsf{P}(v_k):=\textsf{P}_k$. The combination $(\textsf{P}(v_k)+\uptau_1)(\textsf{P}(v_k)+\uptau_2)-\textsf{P}'(v_k)$
is reminiscent of the Miura factorization of a scalar oper. Indeed, in the Gaudin limit the difference operator 
is replaced by a third-order differential operator, while the logarithmic derivative 
\bea
\textsf{P}(u)=\frac{\DP'(u)}{\DP(u)}=\sum_{j=1}^N\frac{1}{u-u_j}\, 
\eea naturally plays a role of the  Miura potential.  This observation suggests that 
the Gaudin limit is governed by
a scalar third-order differential operator
\bea
\la{Diff3}
\mathcal D
=\partial^3+a_1(u)\partial^2+a_2(u)\partial+a_3(u),
\eea
for which the Baxter polynomial $Q_1(u)$ is a solution,
\bea
\mathcal DQ_1(u)=0,
\eea
in agreement with the general oper description of Gaudin models \cite{FrenkelBook,Frenkel2007}.
At this stage, however, $\mathcal D$ should be viewed as a scalar $\gl_3$ oper rather than necessarily as a Miura oper in a fixed factorised gauge.
In (\ref{Diff3}), the coefficients $a_1(u)$ and $a_2(u)$ are 
\bea
\begin{aligned}
a_1(u)&=-2\frac{\DP'(u)}{\DP(u)}-\uptau_1-\uptau_2\, , \\
a_2(u)&=\bigg(\frac{\DP'(u)}{\DP(u)}+\uptau_1\bigg)\bigg(\frac{\DP'(u)}{\DP(u)}+\uptau_2\bigg)-\bigg(\frac{\DP'(u)}{\DP(u)}\bigg)'
+ \sum_{j=1}^N \frac{Q'_1(u_j)}{Q_1(u_j)}\frac{1}{u-u_j}\, .
\end{aligned}
\eea
Consequently, equation (\ref{no_pole}) arises from the condition that $a_3(u)$ has no pole at $u=v_k$. Thus, the non-nested 
Gaudin equations are precisely the pole-free conditions for the scalar $\gl_3$ oper (\ref{Diff3}). 
The relation between the $\mathfrak{sl}_3$ Gaudin model, separation of variables, and scalar third-order  has previously been studied in \cite{Mukhin:2007}.
An important difference is that 
the polynomial \(Q_1(u)\) appearing in the present construction coincides with the
momentum-carrying Baxter polynomial of the nested Bethe ansatz. In the conventional scalar-oper normalization, however, the functions \(\textsf{Q}_i(u)\) are ordered according to
the Miura factorization. Therefore \(Q_1(u)\) should not be identified directly
with the first scalar solution of the oper. Rather, it is expected to be reconstructed from
the corresponding flag data, equivalently from suitable ratios of consecutive
Wronskians of oper solutions.

The explicit analysis presented above was restricted to $\gl_3$, where the leading non-trivial contribution appears at order $\hb^2$. For general $\gl_\ell$, we expect the first non-vanishing contribution to arise at order $\eta^{\ell-1}$, producing a Gaudin equation $\mathcal{R}_\ell^{\rm G}(v_k)=0$. The $\gl_3$ 
example strongly suggests that this equation is equivalent to the requirement that the highest coefficient $a_\ell(u)$ of the associated scalar oper be regular at the Bethe roots.

\section{Conclusions}

In this work we have developed a non-nested formulation of the spectral problem for rational $\mathfrak{gl}_\ell$ spin chains in vector representations. Starting from the $\mathfrak{gl}_3$ and $\mathfrak{gl}_4$ examples, we derived a closed system of equations for the principal Bethe roots by exploiting the associated quantum spectral curve. The construction extends naturally to arbitrary rank and provides a compact and uniform description of the spectrum.


Our derivation follows the general logic of the transfer-matrix characterization based on fusion relations and the associated quantum spectral curve formalism \cite{Maillet:2018bga,Maillet:2019ypa}. The distinguishing feature of the present approach is that the resulting spectral problem is formulated as a closed system of equations for the principal Bethe roots. Once these roots are determined, the corresponding eigenstates can be constructed directly by acting with the $\BB$-operator on the reference state \cite{Gromov:2016itr}. Another attractive feature of the proposed equations is the existence of a well-defined homogeneous limit. 

Another outcome of this work is the analysis of the quasi-classical limit of the non-nested equations. For the $\gl_3$ spin chain, we have shown that the first non-vanishing contribution appears at order $\eta^2$ and gives rise to a Gaudin system whose equations are precisely the regularity conditions for the highest coefficient of an associated scalar third-order oper. 
This establishes a direct link between the non-nested Bethe equations and the oper formulation of the Gaudin model.
It would be interesting to extend this analysis to arbitrary rank. Our results strongly suggest that, for the $\gl_\ell$ spin chain, the first non-trivial contribution appears at order $\eta^{\ell-1}$ and should again admit an interpretation in terms of scalar $\gl_\ell$ opers. Establishing this correspondence in full generality remains an interesting open problem.

A notable feature of the nested Bethe ansatz is the appearance of both momentum-carrying and auxiliary Bethe roots. While the energy and momentum are determined solely by the momentum-carrying roots, the auxiliary roots encode the internal symmetry structure of the states. In this sense, the quantum spectral curve captures not only the spectrum but also the representation-theoretic data associated with the underlying $\gl_\ell$ symmetry in both the twisted and untwisted cases.

An important question is to what extent the auxiliary roots retain physical significance in the thermodynamic limit. Similar questions arise, for instance, in integrable models such as the Gaudin--Yang model \cite{Yang1967,Gaudin1967}, where the nested Bethe ansatz involves momentum-carrying roots together with auxiliary spin roots \cite{GuanLin2023}. Although the energy depends only on the former, the auxiliary roots govern spin degrees of freedom and magnetic observables, reflecting the phenomenon of spin--charge separation. This suggests that the auxiliary sectors appearing in the quantum spectral curve may continue to encode physically meaningful collective degrees of freedom beyond finite volume.

Several directions deserve further investigation. In particular, a systematic thermodynamic analysis of the quantum spectral curve and its continuum limit could shed light on the role of auxiliary roots and the emergence of macroscopic spectral data. It would also be interesting to extend the present construction beyond the vector representation considered here and investigate its counterpart for more general finite-dimensional representations. Another natural question is whether a similar formulation exists for integrable spin chains associated with other simple Lie algebras and their quantum analogues. Furthermore, it would be worthwhile to explore extensions beyond the rational Yangian models studied in this work, including trigonometric spin chains related to quantum affine algebras and, ultimately, elliptic integrable systems. Understanding whether the hierarchical structure 
of transfer matrices and the rank-embedding property
uncovered here persist in these more general settings may provide further insight into the relation between Bethe ansatz techniques, Baxter $Q$-systems, and quantum spectral curves. 



An intriguing open problem is whether the equations $\mathcal{R}_{\ell}=0$ admit a Yang–Yang functional formulation, possibly after an appropriate change of variables or the introduction of integrating factors. Such a formulation could lead to a counterpart of the Gaudin matrix governing the norms of Bethe states. It would also be interesting to clarify the algebraic origin of the recursive hierarchy of the functions $\Phi_\xx$, for example by relating it to a truncated 
$Q$-system, a B\"acklund flow, or another representation-theoretic mechanism underlying the non-nested formulation.
Finally, another interesting question concerns the precise relation 
between the present quantum-spectral-curve construction and the determinant-based approach 
to non-nested Bethe equations developed in \cite{Liashyk:2018qfc}.

We hope that the formulation presented here provides an additional bridge between the nested Bethe ansatz, 
Baxter $Q$-systems, and quantum spectral curves, and serves as a starting point for further developments in the study of higher-rank quantum integrable systems.


\acknowledgments 
G.A and M.G.  are grateful to Paul Ryan for many helpful discussions, while  M.G. also benefited from  discussions with Xiao Wang and Ryo Suzuki.  
We also thank Paul Ryan and Nikita Slavnov for valuable remarks on the manuscript.
G.A. acknowledges support by the DFG under Germany's Excellence Strategy -- EXC 2121 ``Quantum Universe'' -- 390833306 and by the DFG -- SFB 1624 -- ``Higher structures, moduli spaces and integrability'' -- 506632645. H.B. acknowledges Armenian HESC grant 21AG-1C024 for financial support.

\bigskip
\appendix 
\section{Comparison of Nested and Non-nested Bethe Equations}
Here we provide numerical evidence that the nested and non-nested Bethe ansatz equations generate the same set of solutions for the principal Bethe roots. 
As an illustration, we consider the $\gl_3$ 
spin chain of length $N=3$ with $\eta=i$,  twist parameters $\zt=(2,3,5)$, and inhomogeneities $u_1=0$, $u_2=2$, and $u_3=5$. 
The solutions of the standard nested Bethe equations
\bea
\begin{aligned}
&\prod_{\substack{n=1\\ n\neq k}}^{\mb_1}
\frac{v_k-v_n-\eta}{v_k-v_n+\eta}
\prod_{a=1}^{\mb_2}
\frac{v_k-w_a+\eta}{v_k-w_a}
=
-\frac{\zt_1}{\zt_2}
\frac{\DP_1(v_k)}{\DP_2(v_k)},
\qquad k=1,\ldots,\mb_1, \\
&\prod_{k=1}^{\mb_1}
\frac{w_a-v_k}{w_a-v_k-\hb}
\prod_{\substack{b=1\\ b\neq a}}^{\mb_2}
\frac{w_a-w_b-\eta}{w_a-w_b+\eta}
=
-\frac{\zt_2}{\zt_3},
\qquad a=1,\ldots,\mb_2.
\end{aligned}
\eea
for the principal roots $v_k$ are compared with the solutions of the non-nested equations (\ref{alm_B_2}). In both cases, we find  27 physical solutions, 
and the two sets of principal Bethe roots agree exactly, as shown in Table \ref{tab:nested-nonnested-roots}.

The number $\mb_2$ computed for each set of Bethe roots using formula (\ref{AA3}) 
agrees exactly with the value of $\mb_2$ in the nested Bethe table, where the solutions are organized according to the pairs $(\mb_1,\mb_2)$. 
The perfect agreement provides a highly nontrivial consistency check of formula 
(\ref{AA3}), demonstrating that it correctly  reconstructs  the auxiliary 
quantum number $\mb_2$ solely from the principal Bethe  roots.

\begin{table}[ht]
\centering
\tiny
\renewcommand{\arraystretch}{1.15}

\begin{minipage}[t]{0.49\textwidth}
\centering
\begin{tabular}{|c|c|p{0.56\textwidth}|}
\hline
$(m_1,m_2)$ & no. & nested main roots $v$ \\
\hline\hline
$(0,0)$ & 1 & $\{\}$ \\
$(1,0)$ & 1 & $\{2.28641-6.31564 i\}$ \\
$(1,0)$ & 2 & $\{3.87673+0.0579044 i\}$ \\
$(1,0)$ & 3 & $\{0.836854+0.257734 i\}$ \\
$(1,1)$ & 1 & $\{1.90713-1.49292 i\}$ \\
$(1,1)$ & 2 & $\{0.70352-0.110227 i\}$ \\
$(1,1)$ & 3 & $\{4.38935-0.396851 i\}$ \\
$(2,0)$ & 1 & $\{-1.42499-4.19798 i,\;6.13857-4.11766 i\}$ \\
$(2,0)$ & 2 & $\{-0.280816-2.26396 i,\;3.40408+0.321864 i\}$ \\
$(2,0)$ & 3 & $\{1.06693+0.386652 i,\;5.09622-2.12892 i\}$ \\
$(2,1)$ & 1 & $\{0.423333-2.51148 i,\;4.32098-2.52432 i\}$ \\
$(2,1)$ & 2 & $\{1.07446+0.235338 i,\;4.77514-0.967243 i\}$ \\
$(2,1)$ & 3 & $\{1.68804-4.17695 i,\;4.16156-0.424477 i\}$ \\
$(2,1)$ & 4 & $\{0.794711-0.0129991 i,\;2.61137-4.17189 i\}$ \\
$(2,1)$ & 5 & $\{0.118025-1.19373 i,\;3.28806+0.0452867 i\}$ \\
$(2,1)$ & 6 & $\{0.887043+0.0454642 i,\;3.85727-0.342994 i\}$ \\
$(2,2)$ & 1 & $\{0.177795-1.1913 i,\;4.91507-0.968291 i\}$ \\
$(2,2)$ & 2 & $\{1.25323-0.0295194 i,\;5.04325-0.747374 i\}$ \\
$(2,2)$ & 3 & $\{-0.173062-0.813186 i,\;2.78372-0.250332 i\}$ \\
$(3,0)$ & 1 & $\{-2.66535-1.40323 i,\;2.26036-3.16814 i,\;7.40499-1.42863 i\}$ \\
$(3,1)$ & 1 & $\{-1.61122-1.33076 i,\;2.21853-1.95935 i,\;6.3927-1.37656 i\}$ \\
$(3,1)$ & 2 & $\{-1.44406-2.10499 i,\;3.7039-0.325845 i,\;4.74016-2.23584 i\}$ \\
$(3,1)$ & 3 & $\{-0.133782-2.61357 i,\;0.918537+0.0920855 i,\;6.21525-2.14519 i\}$ \\
$(3,2)$ & 1 & $\{-0.93356-0.916263 i,\;2.19784-1.51058 i,\;5.73572-0.906492 i\}$ \\
$(3,2)$ & 2 & $\{-0.388245-0.857632 i,\;3.11883-0.29163 i,\;4.26941-2.18407 i\}$ \\
$(3,2)$ & 3 & $\{0.615365-2.61269 i,\;1.12384+0.0892332 i,\;5.2608-0.809872 i\}$ \\
$(3,3)$ & 1 & $\{-0.673867-0.518017 i,\;2.12257-0.936553 i,\;5.5513-0.54543 i\}$ \\
\hline
\end{tabular}
\end{minipage}
\hfill
\begin{minipage}[t]{0.49\textwidth}
\centering
\begin{tabular}{|c|c|p{0.62\textwidth}|}
\hline
$m_1$ & no. & non-nested main roots $v$ \\
\hline\hline
$0$ & 1 & $\{\}$ \\
$1$ & 1 & $\{1.90713-1.49292 i\}$ \\
$1$ & 2 & $\{2.28641-6.31564 i\}$ \\
$1$ & 3 & $\{0.70352-0.110227 i\}$ \\
$1$ & 4 & $\{3.87673+0.0579044 i\}$ \\
$1$ & 5 & $\{4.38935-0.396851 i\}$ \\
$1$ & 6 & $\{0.836854+0.257734 i\}$ \\
$2$ & 1 & $\{0.423333-2.51148 i,\;4.32098-2.52432 i\}$ \\
$2$ & 2 & $\{-1.42499-4.19798 i,\;6.13857-4.11766 i\}$ \\
$2$ & 3 & $\{1.68804-4.17695 i,\;4.16156-0.424477 i\}$ \\
$2$ & 4 & $\{1.06693+0.386652 i,\;5.09622-2.12892 i\}$ \\
$2$ & 5 & $\{-0.280816-2.26396 i,\;3.40408+0.321864 i\}$ \\
$2$ & 6 & $\{0.794711-0.0129991 i,\;2.61137-4.17189 i\}$ \\
$2$ & 7 & $\{0.118025-1.19373 i,\;3.28806+0.0452867 i\}$ \\
$2$ & 8 & $\{1.07446+0.235338 i,\;4.77514-0.967243 i\}$ \\
$2$ & 9 & $\{0.177795-1.1913 i,\;4.91507-0.968291 i\}$ \\
$2$ & 10 & $\{-0.173062-0.813186 i,\;2.78372-0.250332 i\}$ \\
$2$ & 11 & $\{1.25323-0.0295194 i,\;5.04325-0.747374 i\}$ \\
$2$ & 12 & $\{0.887043+0.0454642 i,\;3.85727-0.342994 i\}$ \\
$3$ & 1 & $\{-1.61122-1.33076 i,\;2.21853-1.95935 i,\;6.3927-1.37656 i\}$ \\
$3$ & 2 & $\{-2.66535-1.40323 i,\;2.26036-3.16814 i,\;7.40499-1.42863 i\}$ \\
$3$ & 3 & $\{-0.133782-2.61357 i,\;0.918537+0.0920855 i,\;6.21525-2.14519 i\}$ \\
$3$ & 4 & $\{-1.44406-2.10499 i,\;3.7039-0.325845 i,\;4.74016-2.23584 i\}$ \\
$3$ & 5 & $\{-0.93356-0.916263 i,\;2.19784-1.51058 i,\;5.73572-0.906492 i\}$ \\
$3$ & 6 & $\{-0.388245-0.857632 i,\;3.11883-0.29163 i,\;4.26941-2.18407 i\}$ \\
$3$ & 7 & $\{0.615365-2.61269 i,\;1.12384+0.0892332 i,\;5.2608-0.809872 i\}$ \\
$3$ & 8 & $\{-0.673867-0.518017 i,\;2.12257-0.936553 i,\;5.5513-0.54543 i\}$ \\
\hline
\end{tabular}
\end{minipage}

\caption{Comparison of the main Bethe roots obtained from the nested and non-nested Bethe equations for the \(\mathfrak{gl}_3\) spin chain of length $N=3$.}
\label{tab:nested-nonnested-roots}
\end{table}

\FloatBarrier

\end{document}